\documentclass[aps,twocolumn,prmaterials,showkeys,showpacs,superscriptaddress]{revtex4-2}

\usepackage[T1]{fontenc}
\usepackage[latin9]{inputenc}
\usepackage{graphicx}
\providecommand{\tabularnewline}{\\}
\usepackage{dcolumn}
\usepackage{bm}
\usepackage{times}
\usepackage{graphics}
\usepackage{float}
\usepackage{caption}
\usepackage{placeins} 
\usepackage{color}
\usepackage{soul}
\usepackage{setspace}
\usepackage{amsmath,amssymb}
\usepackage{ragged2e}
\usepackage{subfig} 
\usepackage{natbib}
\usepackage{comment}
\usepackage{dirtytalk}
\usepackage[dvipsnames]{xcolor}
\usepackage[colorlinks=true,linkcolor=blue,citecolor=blue,urlcolor=blue]{hyperref}

\begin{document}

\title{Tuning Superconductivity in Sputtered W$_\textbf{0.75}$Re$_\textbf{0.25}$ Thin Films}
\author{F.~Colangelo}
\affiliation{Dipartimento di Fisica ``E.R. Caianiello'', Universit\`{a} degli Studi di Salerno, I-84084 Fisciano (Sa), Italy}
\affiliation{CNR-SPIN, c/o Universit\`{a} degli Studi di Salerno, I-84084 Fisciano (Sa), Italy}

\author{F.~Avitabile}
\affiliation{CNR-SPIN, c/o Universit\`{a} degli Studi di Salerno, I-84084 Fisciano (Sa), Italy}

\author{Z.~Makhdoumi~Kakhaki}
\affiliation{Dipartimento di Fisica ``E.R. Caianiello'', Universit\`{a} degli Studi di Salerno, I-84084 Fisciano (Sa), Italy}
\affiliation{CNR-SPIN, c/o Universit\`{a} degli Studi di Salerno, I-84084 Fisciano (Sa), Italy}

\author{A.~Kumar}
\affiliation{Dipartimento di Fisica ``E.R. Caianiello'', Universit\`{a} degli Studi di Salerno, I-84084 Fisciano (Sa), Italy}
\affiliation{CNR-SPIN, c/o Universit\`{a} degli Studi di Salerno, I-84084 Fisciano (Sa), Italy}

\author{A.~Di~Bernardo}
\affiliation{Dipartimento di Fisica ``E.R. Caianiello'', Universit\`{a} degli Studi di Salerno, I-84084 Fisciano (Sa), Italy}

\author{C.~Bernini}
\affiliation{CNR-SPIN Corso Perrone 24, I-16152 Genova, Italy}

\author{A.~Martinelli}
\affiliation{CNR-SPIN Corso Perrone 24, I-16152 Genova, Italy}

\author{A.~Nigro}
\affiliation{Dipartimento di Fisica ``E.R. Caianiello'', Universit\`{a} degli Studi di Salerno, I-84084 Fisciano (Sa), Italy}

\author{C.~Cirillo}
\affiliation{CNR-SPIN, c/o Universit\`{a} degli Studi di Salerno, I-84084 Fisciano (Sa), Italy}

\author{C.~Attanasio}
\affiliation{Dipartimento di Fisica ``E.R. Caianiello'', Universit\`{a} degli Studi di Salerno, I-84084 Fisciano (Sa), Italy}
\affiliation{CNR-SPIN, c/o Universit\`{a} degli Studi di Salerno, I-84084 Fisciano (Sa), Italy}
\affiliation{Centro NANO\_MATES, Universit\`{a} degli Studi di Salerno, I-84084 Fisciano (Sa), Italy}

\begin{abstract}

W$_\text{0.75}$Re$_\text{0.25}$, in its bulk form, has been shown to be an interesting superconducting material due to its multiple crystalline phases, each exhibiting distinct superconducting characteristics. However, little is known about how these phases manifest in thin-film form, where deposition conditions and dimensionality are critical aspects. Here, we investigate superconducting W$_{0.75}$Re$_{0.25}$ thin films deposited via UHV dc magnetron sputtering. In order to tune the crystalline phase of the films, we further explored the effect of incorporating N$_2$ during the deposition. The superconducting and normal-state properties as a function of deposition conditions were investigated, revealing the role of the crystal phase on the film transport properties. 

\end{abstract}

\maketitle

\section{Introduction}

Type--II superconductors with the $\beta$-tungsten ($\beta$-W) structure, also known as A15 compounds (mainly in $A_3B$ composition), represent a technically significant and highly researched group of materials~\cite{Buckel2004}. These systems are distinguished by their high superconducting critical temperatures ($T_c$) and upper critical magnetic fields ($\mu_0H_{c2}$), which have driven their widespread adoption in both theoretical and applied superconductivity research~\cite{Buckel2004}. The A15 compounds exhibit unique structural characteristics, with orthogonally aligned chains of $A$ atoms. This electronic configuration facilitates strong electron-phonon interactions, thus the formation of Cooper pairs, according to BCS theory~\cite{Tinkham}. Such a structure also gives rise to sharp peaks in the density of states, enhancing superconducting performance when the Fermi energy aligns with these peaks~\cite{Buckel2004}. Historically, compounds such as Nb$_3$Ge have showcased remarkable superconducting properties, with Nb$_3$Ge holding the record for the highest $T_c$ for more than a decade~\cite{Gavaler1973,Testardi1974}. Alongside Nb$_3$Ge, materials such as Nb$_3$Sn, and V$_3$Si also play a significant role in technological applications, for instance as superconducting magnets~\cite{Buckel2004,Testardi1967}.

The $\beta$-W structure was observed for the first time in tungsten in 1931~\cite{Hartmann1931}. Typically, tungsten crystallizes in a body centered cubic BCC ($\alpha$-W) structure~\cite{Gibson1964}, which has $T_c \sim 11$~mK. However, thin films of tungsten have shown significantly higher superconducting transition temperatures, ranging between $T_c \sim 2 - 5$~K~\cite{Kammerer1965,Basavaiah1968}. This enhancement has been primarily attributed to a metastable $\beta$-W phase~\cite{Morcom1974,Bond1965}. Additionally, amorphous tungsten ($am$-W) films, with $T_c$ as high as 5~K, can also be obtained by incorporating impurities during the deposition~\cite{Li2008}. However, while superconductivity in $\alpha$-W and $am$-W is a well accepted phenomenon, superconductivity in $\beta$-W remains controversial. In particular, a recent study shows that superconductivity in A15 tungsten is due to an $am$-W layer underneath the $\beta$-W phase, rather than to the $\beta$-W phase itself~\cite{Bagwe2024}. In the case of sputtered thin films, the amorphous $am$-W can be favored by introducing N$_2$ or O$_2$ in the deposition chamber, which has also the effect of stabilizing the $\beta$-W and minimizing the $\alpha$-W phase~\cite{Bagwe2024, Hofer2019, Hofer2023}. In particular, the introduction of N$_2$ in the sputtering process does not necessarily lead to the formation of WN$_x$, since N$_2$ molecules do not bond with W, but they get incorporated inside the film structure as interstitial particles, generating disorder and altering the crystal structure~\cite{Bagwe2024, Hofer2019, Hofer2023}. 

 Superconducting tungsten thin films can be used as a starting point for studying the fundamental properties of W-based superconductors. Among these, W$_{x}$Re$_{1-x}$ is an interesting system, which can crystallize either in an A15, or in a noncentrosymmetric (NCS) structure, which is observed for $x\approx0.25$~\cite{Huang2008,Biswas2011}. NCS materials have garnered significant interest due to the exotic nature of their superconducting order parameter, which can involve a mixture of spin-singlet and spin-triplet components \cite{Bauer2012,Sato2009,Carla2016}. Such systems also hold potential for topologically nontrivial states, making them appealing for spintronics and quantum applications~\cite{Eschrig2015,Manchon2015}. For instance, NCS Nb$_{0.18}$Re$_{0.82}$, which share a similar composition as W$_{0.25}$Re$_{0.75}$ with Nb substituting W, is an interesting material both for fundamental studies and applications. In fact, it has already been shown that Nb$_{0.18}$Re$_{0.82}$ single crystals exhibits two superconducting gaps~\cite{Cirillo2015}, while Nb$_{0.18}$Re$_{0.82}$ thin films are promising for superconducting single-photon detectors (SNSPDs)~\cite{Caputo2017,EsmaeilZadeh2021,Cirillo2024_smspds,Ercolano2023,Ejrnaes2022}. Given the growing interest in W- and Re-based materials for SNSPDs, a systematic study of this composition is timely. {In fact, W-based materials, such as WSi}~\cite{Zhang2016}, {and, more recently, WGe}~\cite{Yang2025} {and W itself,}~\cite{Ma2025} {showed high efficiency at long wavelengths, while Re-based materials, as NbRe and NbReN,} {have recently showed single-photon sensitivity up to 2$\mu$m}~\cite{Cirillo2024_smspds}. On the other hand, the A15 structure is observed in  W$_{0.75}$Re$_{0.25}$. As pure tungsten, bulk W$_{0.75}$Re$_{0.25}$ can crystallize both in the BCC $\alpha$-phase ($\alpha$-W$_{0.75}$Re$_{0.25}$), and in the A15 crystal structure ($\beta$-W$_{0.75}$Re$_{0.25}$)~\cite{Federer1965, Easton1974}. As in tungsten, when the $\beta$-phase is observed, W$_{0.75}$Re$_{0.25}$ exhibits a larger $T_c$. In particular, bulk $\beta$-W$_{0.75}$Re$_{0.25}$ shows a $T_c$ around 10~K~\cite{Easton1974}, while $\alpha$-W$_{0.75}$Re$_{0.25}$ has a $T_c\sim$~5~K~\cite{Easton1974}. These similarities with the pure tungsten might be due to their common crystal structure, since, in W$_{0.75}$Re$_{0.25}$, Re substitutes some W atoms in the tungsten lattice, without altering the overall structure of tungsten~\cite{Easton1974}. However, despite the promising superconducting behavior of W$_{0.75}$Re$_{0.25}$ in bulk form~\cite{Federer1965, Easton1974}, its thin-film properties remain largely unexplored.
 
{This paper focuses on the growth and characterization of W$_{0.75}$Re$_{0.25}$ thin films, aiming to investigate the fundamental physics of this material and assess its potential for superconducting applications. In particular, we sputtered W$_{0.75}$Re$_{0.25}$ films with varying thicknesses and N$_2$ concentrations during the deposition process and investigated how these parameters affect their structural and transport properties to establish a first comprehensive picture of their normal- and superconducting-state behavior.}

\section{Experiment}

W$_{0.75}$Re$_{0.25}$ films were sputtered using an ultra-high (UHV) vacuum direct current (DC) magnetron system. A stoichiometric W$_{0.75}$Re$_{0.25}$ target from Testbourne (99.99\% purity) with a diameter of 5~cm and a thickness of 3~mm was used, {and the target to substrate distance was kept in the range 15-20~cm.} The system's base pressure was maintained in the low 10$^{-8}$~mbar range. Precise control over Ar and N$_\text{2}$ gas flux was achieved by two separate mass flow controllers. A series of sputtering depositions were performed to identify the optimal growth conditions. Different substrates and sputtering powers were also tested, with negligible impact on the overall quality of the samples. Here, we focus on films deposited on Si(100) substrates at a sputtering power of 150~W and room temperature. Films were produced, ranging in thickness from 3 to 100~nm, at a fixed Ar gas pressure of $P_{\mathrm{Ar}}=3.0$~$\mu$bar. A stable deposition rate of 0.16~nm/s was monitored using a quartz crystal microbalance calibrated with a Bruker DektakXT profiler.

In order to tune the different crystalline phases, and therefore the normal- and superconducting-state properties of the films, N$_2$ gas was introduced into the deposition chamber. The N$_2$ flux was controlled as a percentage of the total gas mixture (Ar and N$_2$) incoming flux, with depositions conducted at two different N$_2$ concentrations (5 and 7.5\%) and thicknesses ($d=25,\,40,\,60$~nm). For clarity, each sample deposited in pure Ar atmosphere is labeled with the letter P, preceded by a number denoting film thickness in nanometers. Samples deposited in an Ar/N$_2$ mixture are denoted by an N (standing for nitrogen) prefix, followed by the N$_2$ flux percentage used during deposition. For instance, sample 25P is the film of 25~nm deposited in an Ar atmosphere, while sample 25N7.5 is the 25~nm thick film deposited in an Ar/N$_2$ mixture with the N$_2$ being the 7.5\% of the total incoming gas flow. {Additionally, samples with 10\% N$_2$ incoming flow were also deposited and their analysis is reported in the Supplemental Material.}

Phase analysis of the films was performed by Scanning Electron Microscopy (Zeiss GeminiSEM 360), equipped
with an EDS (Energy-Dispersive X-ray Spectroscopy) microanalysis probe (Oxford X-Max 20) for
quantitative elemental analysis. The analysis was conducted using a beam accelerating voltage of $5$~keV and a specimen current of $1.14$~nA. The crystalline properties of these films were characterized with a $\theta-2\theta$ BRUKER D2 X-Ray Diffractometer (XRD) system equipped with monochromatic CuK$\alpha_1$ radiation ($\lambda = 1.5406$~\AA). Electrical transport measurements were conducted in a Cryogen-Free High Field (7~T) Measurement System by CRYOGENIC, Ltd. Resistances measurements have been acquired with a power supply Keithley 6121 operating together a nanovoltmeter Keithley 2182 in Delta mode, in a standard four-wire configuration on unstructured samples with an excitation current of 10~$\mu$A. Resistivity ($\rho$) was determined using the van der Pauw (vdP) method ~\cite{Pauw1958,Koon1992}. 

\section{Results}
\subsection{EDS and XRD characterization}

The chemical composition of the films was systematically measured by energy dispersive spectroscopy to check both the stoichiometry of the pure films and the nitrogen content of the those deposited in the Ar/N$_\text{2}$ mixture. Table~\ref{EDS} {lists the chemical composition of the inspected films, obtained by averaging the results of approximately ten different analysis points. The elements are reported as their percentage over the whole films' composition, which results close to the target's nominal stoichiometry. Additionally, although affected by oxidation in air, these data demonstrate the effectiveness of the applied nitriding process.}

\setlength{\tabcolsep}{5pt}
\renewcommand{\arraystretch}{1.5}
\begin{table}[!h]
\vspace{3mm}
\begin{centering}
\caption{Chemical composition of the inspected thin films and their standard variation as obtained by EDS analysis.}
\label{EDS}
\begin{tabular}{c  c  c  c  c}
\hline
\hline
Series & W(\%) & Re(\%) & N(\%)  \tabularnewline
\hline 
-P & $78.5\pm0.2$ & $21.5\pm0.2$ & / \tabularnewline

-N5 & $72.7\pm0.3$ & $21.7\pm0.3$ & $5.6\pm0.5$\tabularnewline

-N7.5 & $68.9\pm0.3$ & $22.1\pm0.4$ & $9.0\pm0.7$\tabularnewline

\hline  
\hline
\end{tabular}
\par\end{centering}
\end{table}

\begin{figure}[h]
\centerline{\includegraphics[width=9.5cm]{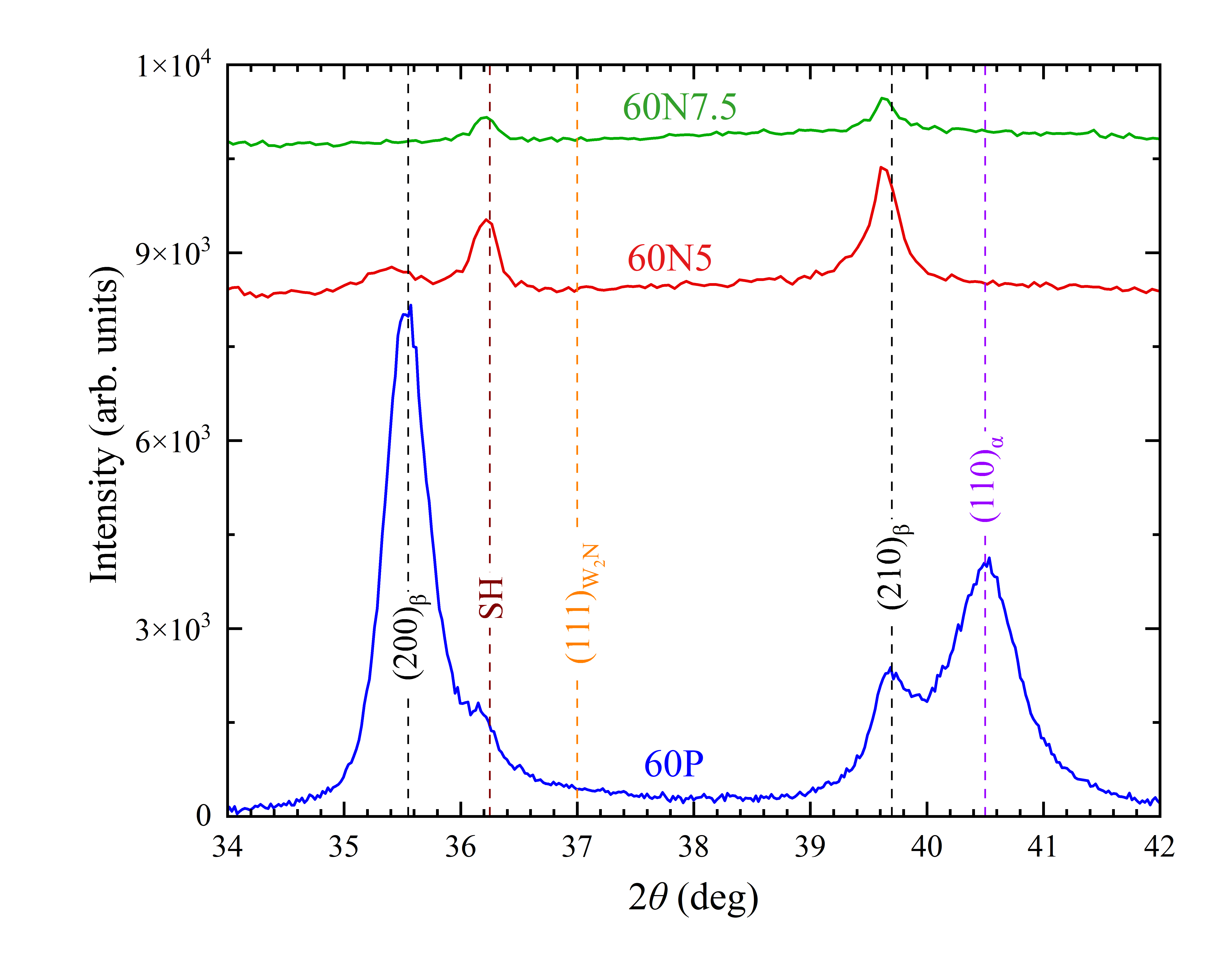}}
\caption{XRD data for W$_{0.75}$Re$_{0.25}$ 60-nm thin films grown on Si substrates. The data of samples 60P, 60N5 and 60N7.5 are reported with a blue, red and green line, respectively. The $\left(200\right)_\beta$ and $\left(210\right)_\beta$ peak of $\beta$-WRe are indicated by dashed black lines, while the $\alpha$-phase peak $\left(110\right)_\alpha$ is indicated with a purple dashed line. The peaks of the sample holder (SH) and the $\left(111\right)$ peak of W$_2$N (absent) are indicated by dashed brown and orange lines, respectively.} 
\label{XRD}
\end{figure}
\begin{figure*}
	\centering
	\subfloat
	{\includegraphics[width=.5\textwidth]{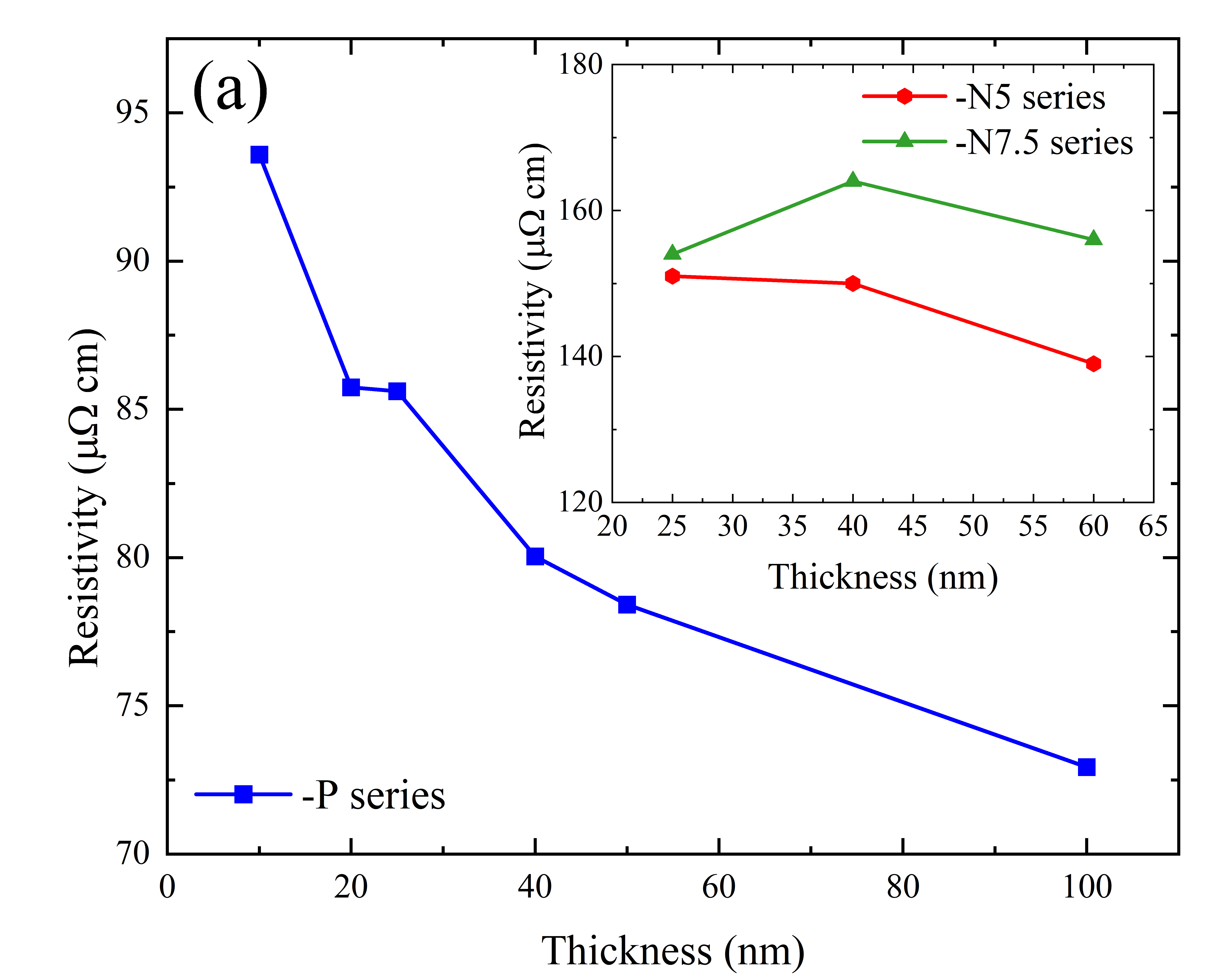}} 
	{\includegraphics[width=.505\textwidth]{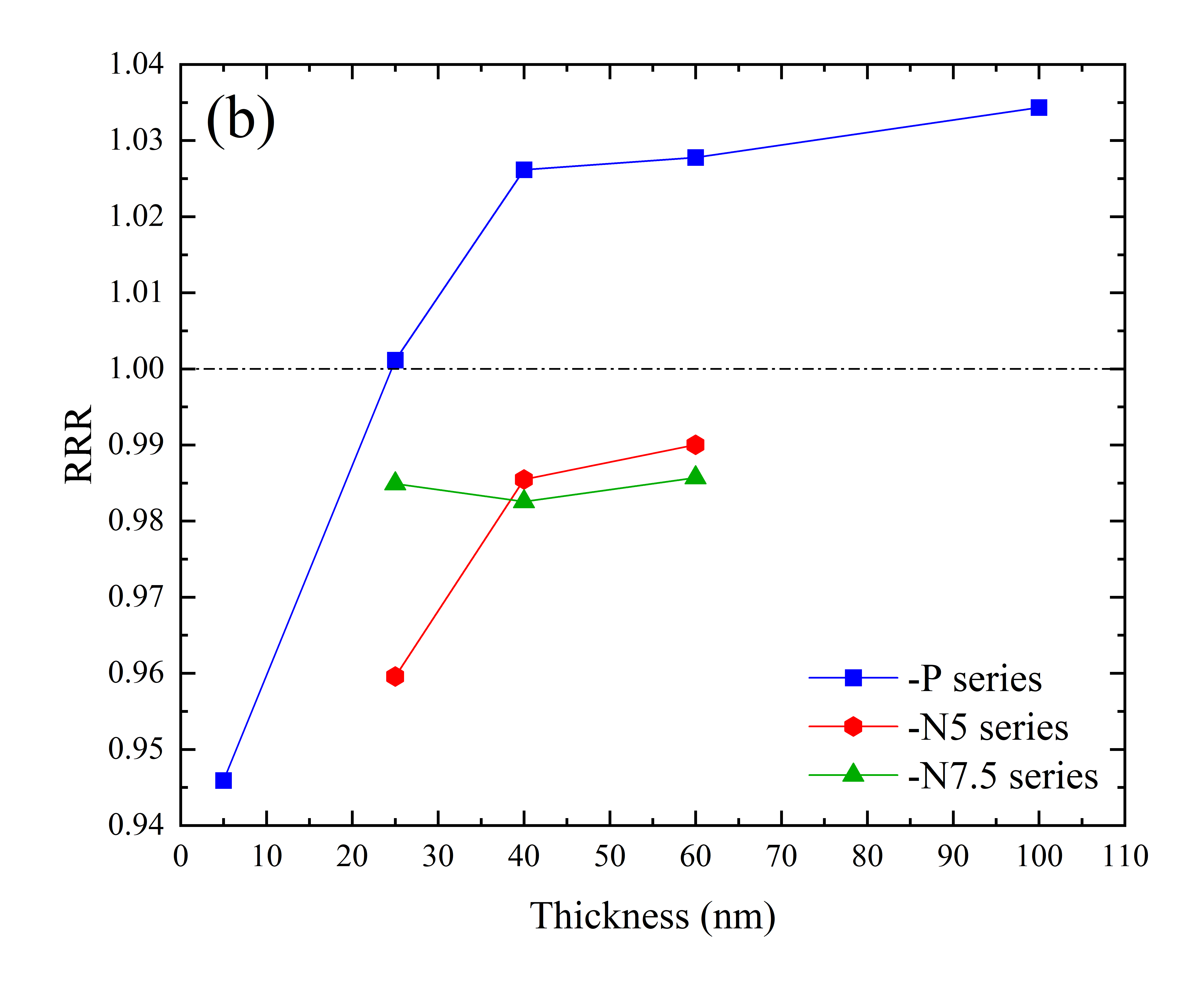}}
        \caption{(a) $\rho(d)$ at 10~K of the -P series, with the inset showing the trend for the -N5, and -N7.5 series; (b) $RRR$ as a function of $d$ for the four samples series. The dashed line corresponds to $RRR=1$.}
	\label{norm}
\end{figure*}

For the films of the -P series, the XRD analysis revealed the presence of several peaks and the coexistence of the $\alpha$- and $\beta$-WRe phases. In Fig.~\textcolor{blue}{\ref{XRD}}, a high-angle $\theta$-2$\theta$ scan performed on sample 60P is reported by a blue line. The different peaks corresponding to diffraction planes were labeled using XRD data from previous studies on bulk W$_{x}$Re$_{1-x}$ and W thin films. In particular, from Refs.~\cite{Federer1965, Easton1974} it was possible to identify the $\beta$-WRe peaks $\left(200\right)_\beta$ and $\left(210\right)_\beta$, while Refs.~\cite{Bagwe2024, Holzman2019} were used to label the $\left(110\right)_\alpha$ peak of the $\alpha$-WRe. Additionally, a minor sample holder peak (SH in Fig.~\textcolor{blue}{\ref{XRD}}) is observable in the tail of the $\left(200\right)_\beta$ peak. 
 
The $\beta$-WRe lattice parameter $a_\beta$ was evaluated from the Bragg law, leading to $a_\beta=5.05\pm0.01$~\AA. This result aligns well with the bulk-W$_\text{0.78}$Re$_\text{0.22}$ $\beta$-phase lattice parameter, $a_\beta^\text{bulk}=5.0182\pm0.0005$~\AA, reported in Ref.~\cite{Federer1965}. Using the Miller indices reported in Refs.~\cite{Bagwe2024, Holzman2019} for the $\left(110\right)_\alpha$ peak, the $\alpha$-WRe crystal side $a_\alpha$ results to be $a_\alpha=3.15\pm0.01$~\AA. Despite the lack of data in literature regarding W$_\text{0.75}$Re$_\text{0.25}$, the obtained $a_\alpha$ well-match with the values $a_\alpha^\text{bulk}=3.165\pm0.001$ and $a_\alpha^\text{bulk}\approx3.15$ reported for bulk W$_\text{0.87}$Re$_\text{0.13}$~\cite{Federer1965} and W$_\text{0.73}$Re$_\text{0.27}$~\cite{Easton1974}, respectively.

{The nitrogen concentrations of 5\% and 7.5\% were selected to systematically explore the structural evolution of W$_{0.75}$Re$_{0.25}$ films as a function of N$_2$ incorporation. This range spans from low to moderately high nitrogen levels. The XRD patterns for samples 60N5 and 60N7.5 are also shown in Fig.}~\textcolor{blue}{\ref{XRD}}, {and they display noticeable differences from the 60P spectrum. All XRD measurements were performed under identical experimental conditions, including scan geometry and step time. The resulting profiles are presented without normalization, in order to preserve the absolute intensity of the diffraction peaks. This approach facilitates direct comparison of relative crystallinity across the samples and is consistent with previous reports on nitrogen incorporation in tungsten-based films}~\cite{Bagwe2024, Hofer2019, Hofer2023}. {The curves have been vertically shifted for clarity. Importantly, the $\left(110\right)_\alpha$ peak of the $\alpha$-WRe phase is absent in both nitrogen-doped samples, already at 5\% N$_2$, indicating a significant structural change. Indeed, this result confirms that N$_2$ stabilizes the $\beta$-WRe structure. The intensity of the $\left(200\right)_\beta$ and $\left(210\right)_\beta$ peaks is strongly reduced in 60N5 and barely detectable in 60N7.5. Additionally, their positions is slightly shifted to the left, indication an increase of the cell size. It is also worth noting the absence of the W$_2$N $\left(111\right)$ peak (orange dashed line in Fig.}~\textcolor{blue}{\ref{XRD}}), {which suggests that nitrogen dissolves within the $\beta$-WRe structure forming a solid solution, in agreement with previous studies}~\cite{Bagwe2024, Holzman2019}. {Increasing the nitrogen content to 7.5\% further reduces the intensity of the diffraction peaks, suggesting enhanced amorphization. Although no intermediate nitrogen concentrations were investigated in this study, the clear and distinct trend observed across the selected values supports the choice as representative of different structural regimes. Future work will expand this parameter space to include finer increments of N$_2$ concentration. The role of nitrogen in modifying the crystallinity of W-based films remains complex and has been reported inconsistently in the literature, with both amorphization and crystallization trends observed}~\cite{Bagwe2024, Hofer2023}, {underscoring the need for further systematic investigations. It is also noted that the peak near $2\theta \approx 36.2^\circ$ shows variable intensity across the samples. While this feature corresponds to a sample holder reflection, it may partially overlap with the (200)$_\beta$ diffraction peak of the $\beta$-WRe phase. This overlap, as well as possible differences in the alignment, complicates its interpretation.}

\begin{figure*}
	\centering
	\subfloat
	{\includegraphics[width=.52\textwidth]{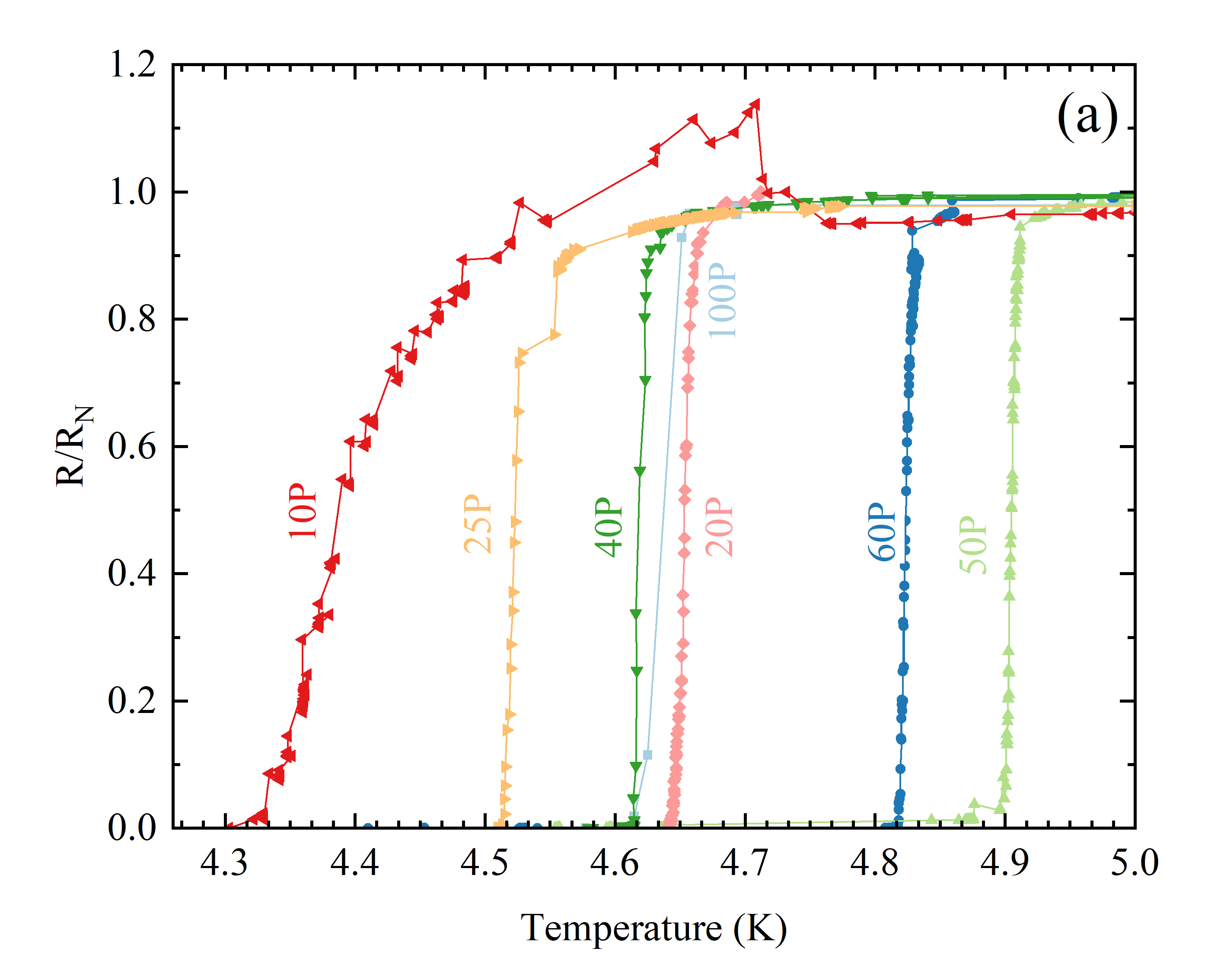}} 
	{\includegraphics[width=.56\textwidth]{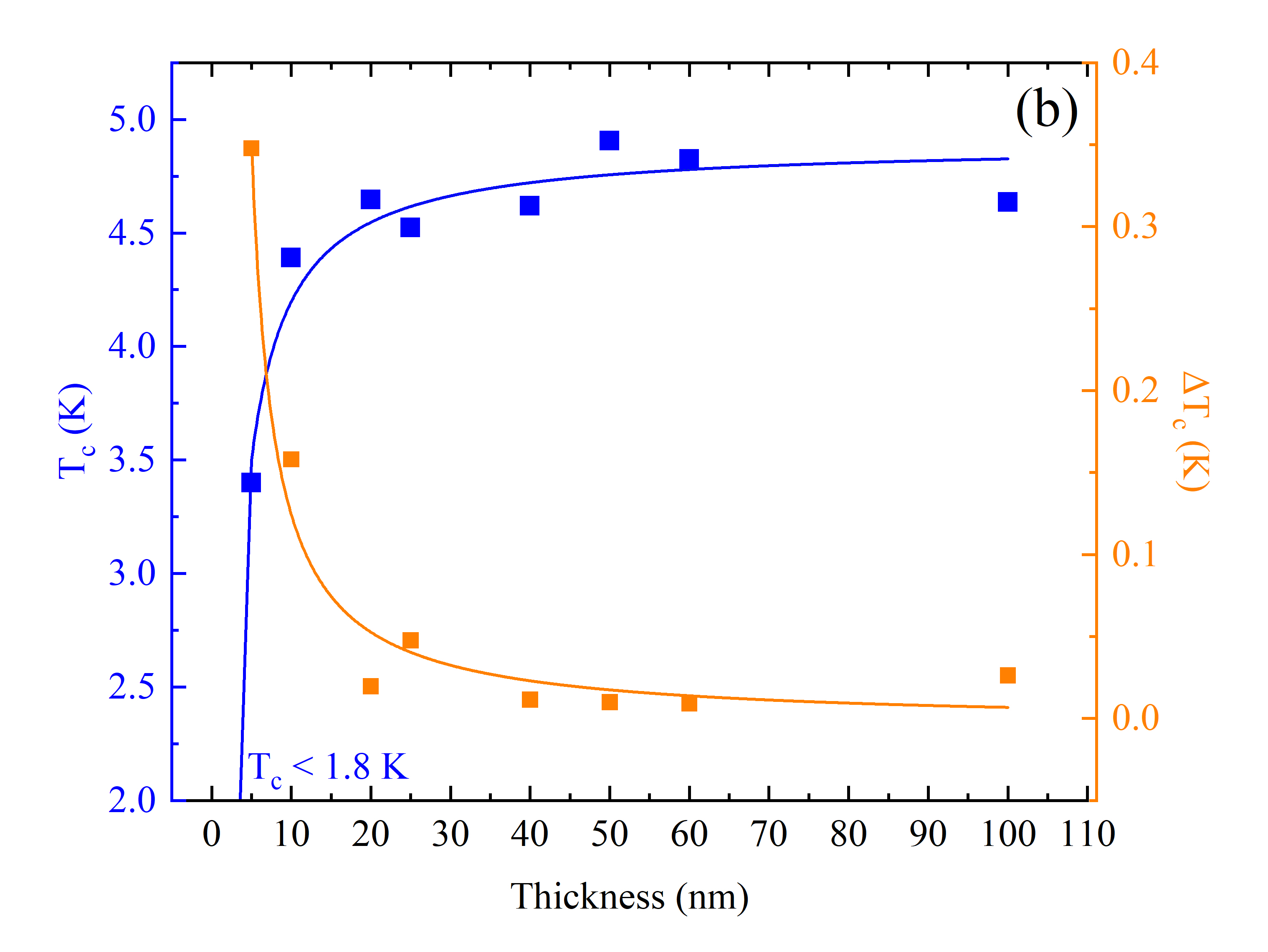}}
        \caption{(a) Normalized resistive transitions of -P series samples. (b) $T_c$ and $\Delta T_c$ as functions of $d$ for the same films. No superconducting transition has been observed in sample 3P down to 1.8~K. The solid blue and orange lines are guides for the eye for the $T_c(d)$ and $\Delta T_c(d)$ data, respectively.}
	\label{series}
\end{figure*}

\subsection{Normal-state properties}

Fig.~\textcolor{blue}{\ref{norm}(a)} shows the $\rho$ versus $d$ dependence at 10~K for the films of the -P series. The increasing $\rho$ behavior at low thickness, typical of metallic thin films, can be attributed to the dominance of surface scattering and reduced grain size, which results in higher resistivity due to increased electron scattering at the boundaries~\cite{Fuchs1938,Sondheimer1952,Haq1982,Andreone1995}. The $\rho(d)$ trend of the samples deposited with an Ar/N$_2$ mixture are reported in the inset in Fig.~\textcolor{blue}{\ref{norm}(a)}. The films of the -N series exhibit a larger resistivity compared to those of the -P series. In particular, $\rho$ scales with the nitrogen content with the -N7.5 series exhibiting larger resistivity compared to -N5 one. Moreover, for both series $\rho$ appears to be almost thickness independent. In fact, due to their poor crystallinity, the influence of the grain boundary is reduced. {Similar results have also been observed in the case of pure W thin films, as reported in Ref.}~\cite{Bagwe2024}.

The Residual-Resistance Ratio (RRR) was calculated as the ratio between $\rho^\text{250K}\equiv\rho(T=250\text{~K})$ and $\rho^\text{10K}\equiv\rho(T=10\text{~K})$, and it is presented in Fig.~\textcolor{blue}{\ref{norm}(b)} as a function of $d$. All the samples have shown an RRR close to 1, as typically found in dirty superconductors, such as NbN~\cite{Nigro1988,Marsili2009}, NbRe~\cite{Carla2016}, and NbReN~\cite{Cirillo2021}, as well as in other W-based superconductors, for instance WSi~\cite{Zhang2016}. The RRR values exhibit a decreasing trend as thickness is reduced for all series, except for the -N7.5 series, which appears thickness-independent. {The variation of RRR with thickness and nitrogen content can be understood by considering the presence of an amorphous interfacial layer at the film/substrate boundary, as reported in Ref.}~\cite{Bagwe2024}. {In the -P series, RRR decreases with decreasing thickness, suggesting that the disordered interface contributes increasingly to charge scattering in thinner films. In contrast, for the -N7.5 films, which are nearly amorphous throughout, RRR appears independent on thickness, consistently with uniform disorder across the film. These trends support a model where amorphous or disordered regions reduce the temperature dependence of $\rho$ and suppress RRR.} This behavior is therefore in agreement with the poor crystallinity of the -N7.5 series compared to the others, as shown in the XRD analysis. Additionally, the overall trend reveals that the higher crystallinity of a series corresponds to a larger RRR. 

\subsection{Superconducting critical temperatures}

Fig.~\textcolor{blue}{\ref{series}(a)} shows the resistance versus temperature, $R(T)$, curves for the -P series samples at various thicknesses, normalized by the normal state value, $R_N$, at the onset of the transition. Fig.~\textcolor{blue}{\ref{series}(b)} reports both $T_c$ and the transition width $\left(\Delta T_c\right)$ as functions of $d$. $T_c$ is defined as the temperature at which the resistance becomes 50\% of $R_N$, while $\Delta T_c$ is defined as the difference between the temperatures at which the resistances are 90\% and 10\% of their onset values. {This analysis is relevant both for understanding finite-size effects in disordered superconductors and for optimizing WRe-based films in device applications requiring thin and stable superconducting layers.} For $d>10$~nm, the transitions are sharp and both $T_c$ and $\Delta T_c$ are almost independent on $d$, with saturation values of $T_c\simeq4.7$~K and $\Delta T_c\simeq20$~mK. Below this threshold, $T_c$ decreases with $d$, while $\Delta T_c$ becomes larger, as expected for metallic superconductors~\cite{Simonin1986}. Typically, a dependence of $T_c$ and $\Delta T_c$ on $d$ is expected also for larger thicknesses. However, as further discussed later in the text, this can be explained by considering that superconductivity in our WRe thin films might be only due to a certain amorphous layer thinner than the actual $d$.

\begin{figure}
    \subfloat
    {\centerline{\includegraphics[width=9.2cm]{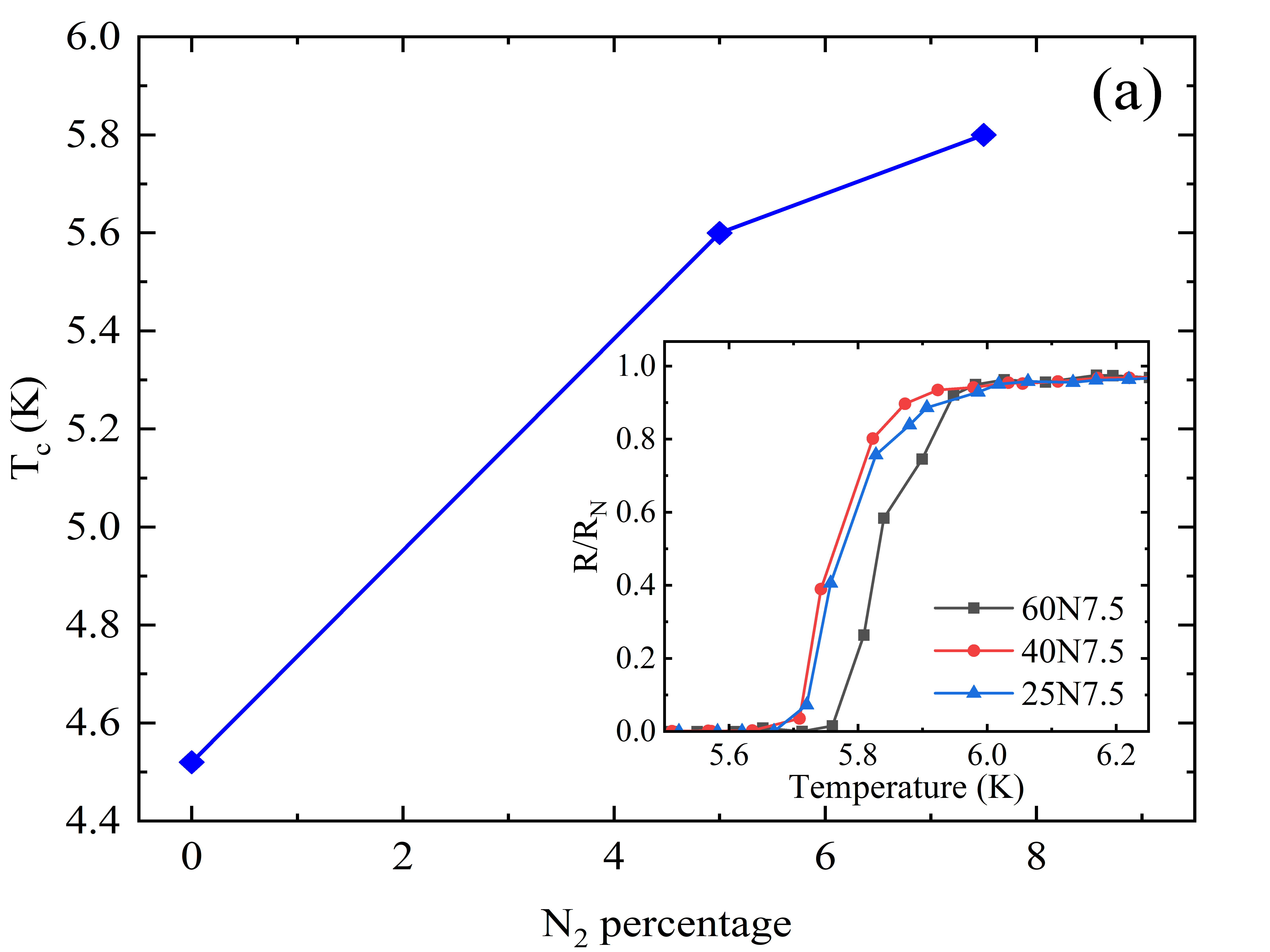}}}\\
    {\centerline{\includegraphics[width=9.2cm]{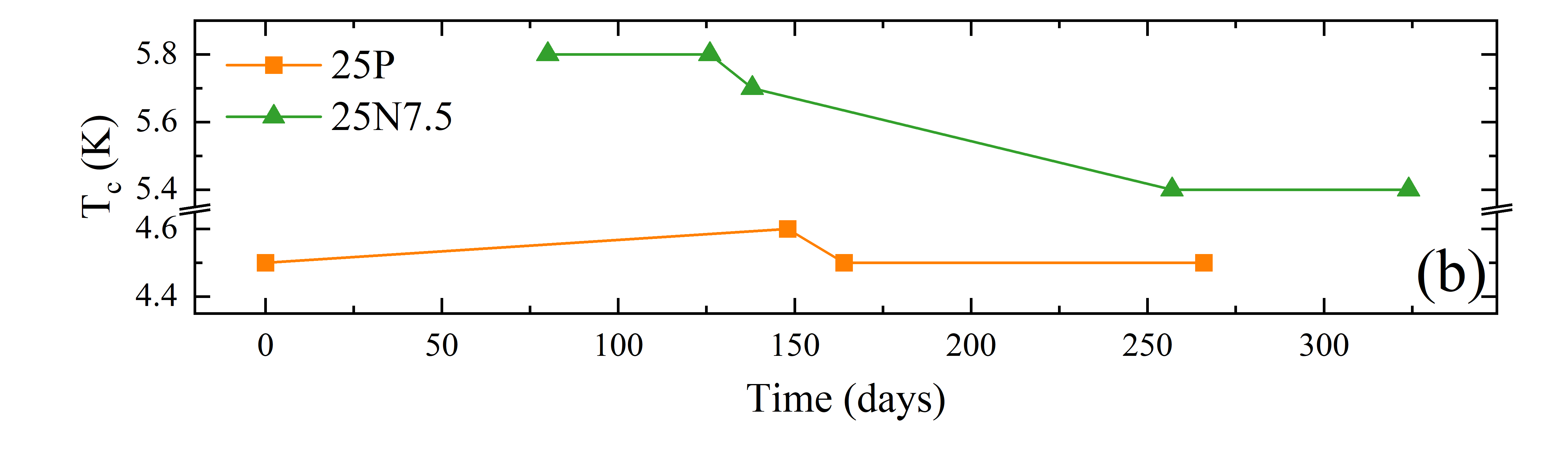}}}
    \caption{(a) $T_c$ of the 25-nm films as a function of the N$_2$ flux percentage, with the inset showing the normalized $R(T)$ for the -N7.5 series samples; (b) $T_c$ over time for the 25P and 25N7.5 films.} 
\label{Tc_N}
\end{figure}

The -N~series samples exhibit larger $T_c$s with respect to those of the -P series. In particular, at $d=25$~nm, the film with the largest $T_c$ is 25N7.5, as reported in Fig.~\textcolor{blue}{\ref{Tc_N}(a)}, where the $T_c$ for samples with fixed thickness ($d=25$~nm) but grown at different N$_2$ gas flux percentages are reported. The $T_c$ of -N7.5 series appears almost thickness-independent for the tested samples, similarly to what was previously observed for the -P series for $d>10$~nm. {The superconducting resistive transitions of the -N7.5 series films are reported in the inset in Fig.}~\textcolor{blue}{\ref{Tc_N}(a)}. It is worth noting that the samples display an excellent stability of $T_c$ over time, despite being kept in air and undergoing many temperature cycles, as reported in Fig.~\textcolor{blue}{\ref{Tc_N}(b)}. Measurements over a time period of more than 8 months show a $T_c$ time stability for the film 25P, with oscillation of the order of $0.1$~K. On the other hand, $T_c$ of the sample 25N7.5 has decreased by $0.4$~K in approximately 6 months. {These results demonstrate the robustness of the superconducting phase under ambient conditions. Similar stability was observed in several other films over similar intervals, confirming the reproducibility of this behavior across both series.}

\begin{figure}
	\centerline{\includegraphics[width=0.5\textwidth]{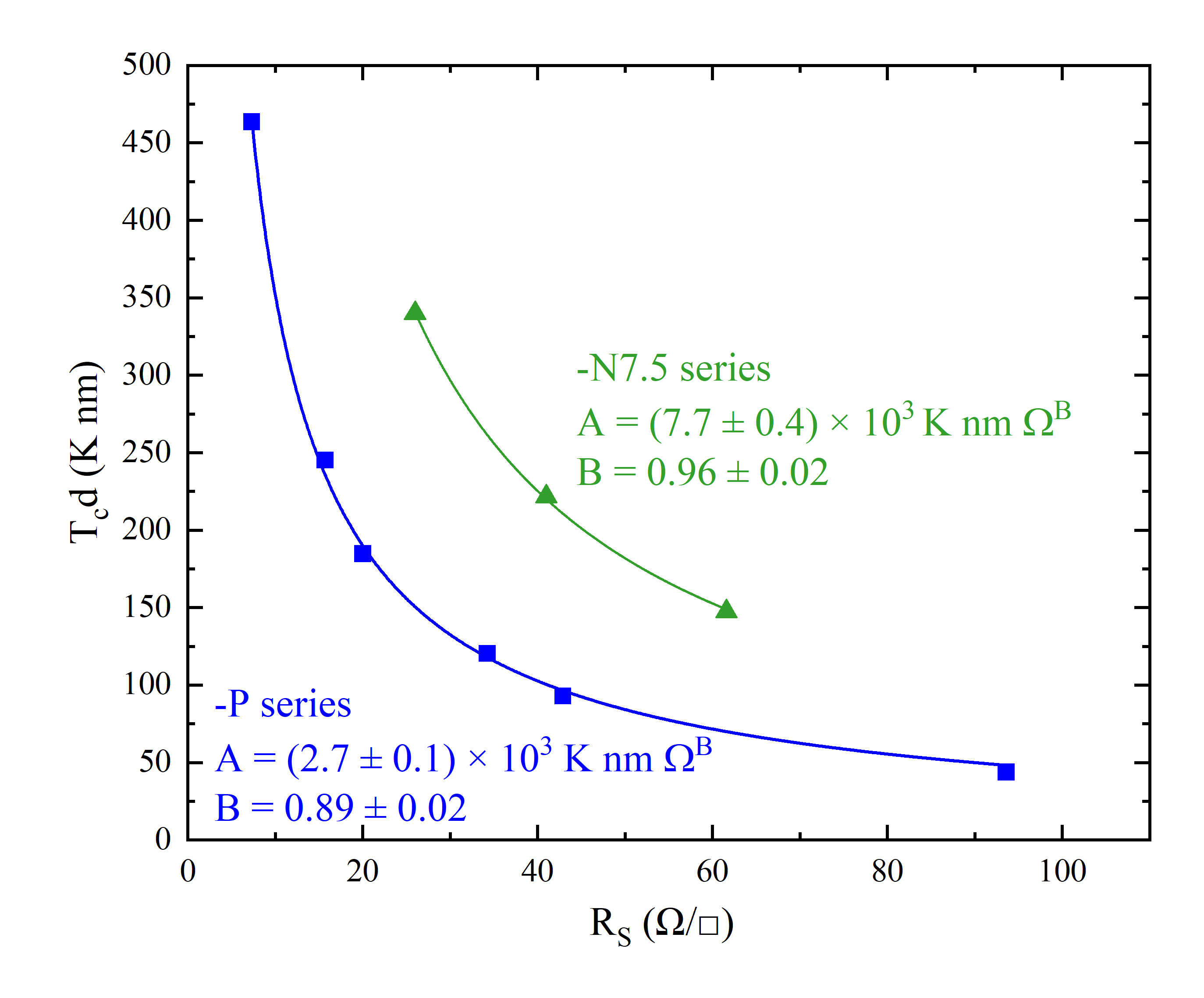}} 
        \caption{Fitting of the relation $d\cdot T_c = A\,R_S^{-B}$ for samples of the -P (in blue) and the -N7.5 (in green) series. The values of the fitting parameters $A$ and $B$ are reported in the panel using the same color scheme.}
	\label{dTc}
\end{figure}

We further analyze the scaling of $T_c$ according to the model discussed in Ref.~\cite{Ivry2014}. This approach examines the product $d\cdot T_c$ as a function of the sheet resistance $R_S$, where $R_S = \rho/d$~\cite{Ivry2014}. This analysis provides insights into the interaction between superconductivity and disorder in materials. Additionally, it is particularly interesting since $T_c$ seems thickness independent above 10~nm.  It has been shown that $d\cdot T_c$ consistently follows the same functional dependence on $R_S$ across a wide range of materials. Data from over 30 materials reported in the past five decades indicate that $d\cdot T_c$ scales with $R_S$ as $d\cdot T_c = A\,R_S^{-B}$, where $A$ and $B$ are fitting parameters~\cite{Ivry2014}. These parameters are not independent, as they satisfy the relationship $A \sim e^{B}$. In particular, $B$ has been found to lie within the range $0.2-1.9$, with a mean value of $0.95$~\cite{Ivry2014}. In Fig.~\textcolor{blue}{\ref{dTc}}, we present the fits of the data for the films of -P and -N7.5 series using this model. The fitting parameters obtained for the -P series are $A=(2.7\pm0.1)\times10^3\,\text{K nm }\Omega^B$ and $B=0.89\pm0.02$, while for the films of the -N7.5 series, the parameters are $A=(7.7\pm0.4)\times10^3\,\text{K nm }\Omega^{B}$ and $B=0.96\pm0.02$. These values are consistent with the general trend observed in Ref.~\cite{Ivry2014} for other materials. In particular, the fitting parameters $A$ and $B$ are usually larger for more disordered films, which is also the case of our W$_\text{0.75}$Re$_\text{0.25}$ films, with the -N7.5 series showing larger fitting parameters compared to -P series. Moreover, $A$ and $B$ of our films are lower than those reported for W$_{0.30}$Re$_{0.70}$, which are $A=1.5\pm0.1\times10^3\,\text{K nm }\Omega^B$ and $B=1.08$~\cite{Ivry2014}. 

\subsection{$\mu_0H_{c2}$ phase diagrams}

\begin{figure*}
	\centering
        {\includegraphics[width=1\textwidth]{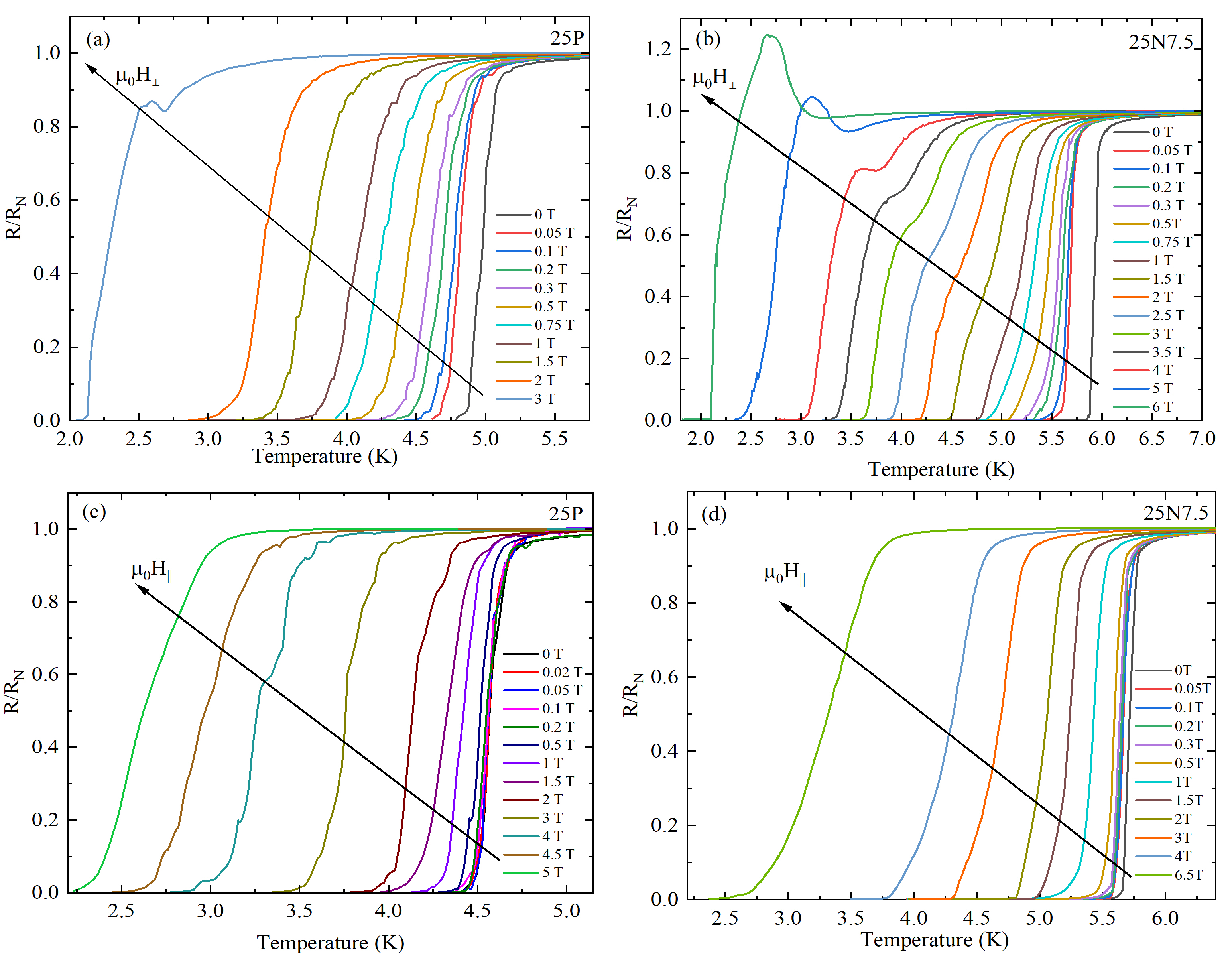}} 
        \caption{Normalized R(T) curves of samples 25P in (a) perpendicular and (c) parallel magnetic fields. Panels (b) and (d) shows the normalized resistive transitions of sample 25N7.5 in perpendicular and parallel fields, respectively.}
	\label{RHT}
\end{figure*}

Electric transport measurements in magnetic field ($\mu_0H$) were conducted for both the -P and -N7.5 series. Based on the previous observations, sample 25N7.5 was chosen as the most significant among films deposited in an Ar/N$_2$ mixture, since it has the largest $T_c$ and the poorest crystallinity,

Fig.~\textcolor{blue}{\ref{RHT}} displays the $R(T)$ curves under perpendicular and parallel magnetic field, for 25P and 25N7.5, in panels (a, c) and (b, d), respectively. The slight difference in $T_c$ at zero field between the parallel and perpendicular field configurations arises from the differing thermal coupling of the sample to the temperature sensor in the two cases. The superconducting transition broadens when the applied magnetic field increases. Moreover, panels (a) and (b) reveal a field-driven peak effect. For sample 25N7.5, this effect starts as a shoulder around $\mu_0H\approx2.5$~T, developing into a distinct peak as the field increases, even surpassing the normal-state resistance above $\mu_0H\approx5$~T. Notably, this peak effect is not observed when the field is applied parallel to the film surface. A similar phenomenon occurs in the $R(T)$ curve of the 25P sample; however, due to its lower $T_c$, measurements could not be extended beyond $3$~T for perpendicular fields. In the literature, peak effects are often observed in $R(T)$, although they are typically suppressed by applying a magnetic field~\cite{Lindqvist1990,Nordstrom1992,Santhanam1991,Zhang2013}. This occurrence is typically associated with the formation of superconductor-normal metal-superconductor (SNS) junctions. In this case, the resistance peak is maximum in zero field condition. However, in our W$_{0.75}$R$_{0.25}$ thin films, the peaks only appear when a certain magnetic field is applied, and their magnitude increases with the field. Field-activated peaks have been reported in studies of high-temperature superconductors (HTS)~\cite{He1990,Suzuki1994,Sambandamurthy2005} where they are attributed to other phenomena. Additionally, field-actived peaks have also been observed in Nb/SR-STO films~\cite{Singh2020}, because of perpendicular components of the bias current. To investigate whether the peak effect observed here could be an artifact of the vdP contact configuration~\cite{Vaglio1993}, the 25N7.5 sample was also measured in an inline contact configuration. In this setup, no peak effect was observed. As shown in Ref.~\cite{Vaglio1993}, a vdP resistance peak occurs if a current-voltage contact pair reaches zero resistance before the others. This effect disappears in inline contact measurements. However, Ref.~\cite{Vaglio1993} examines this effect in zero field, whereas here the peak effect is activated by increasing $\mu_0H$. In the Supplemental Material, an extended version of the Ref.~\cite{Vaglio1993} model is discussed, to reproduce the peak effect observed in the resistive transitions of sample 25N7.5 in perpendicular field.

\begin{figure}
    \subfloat
    {\centerline{\includegraphics[width=9.5cm]{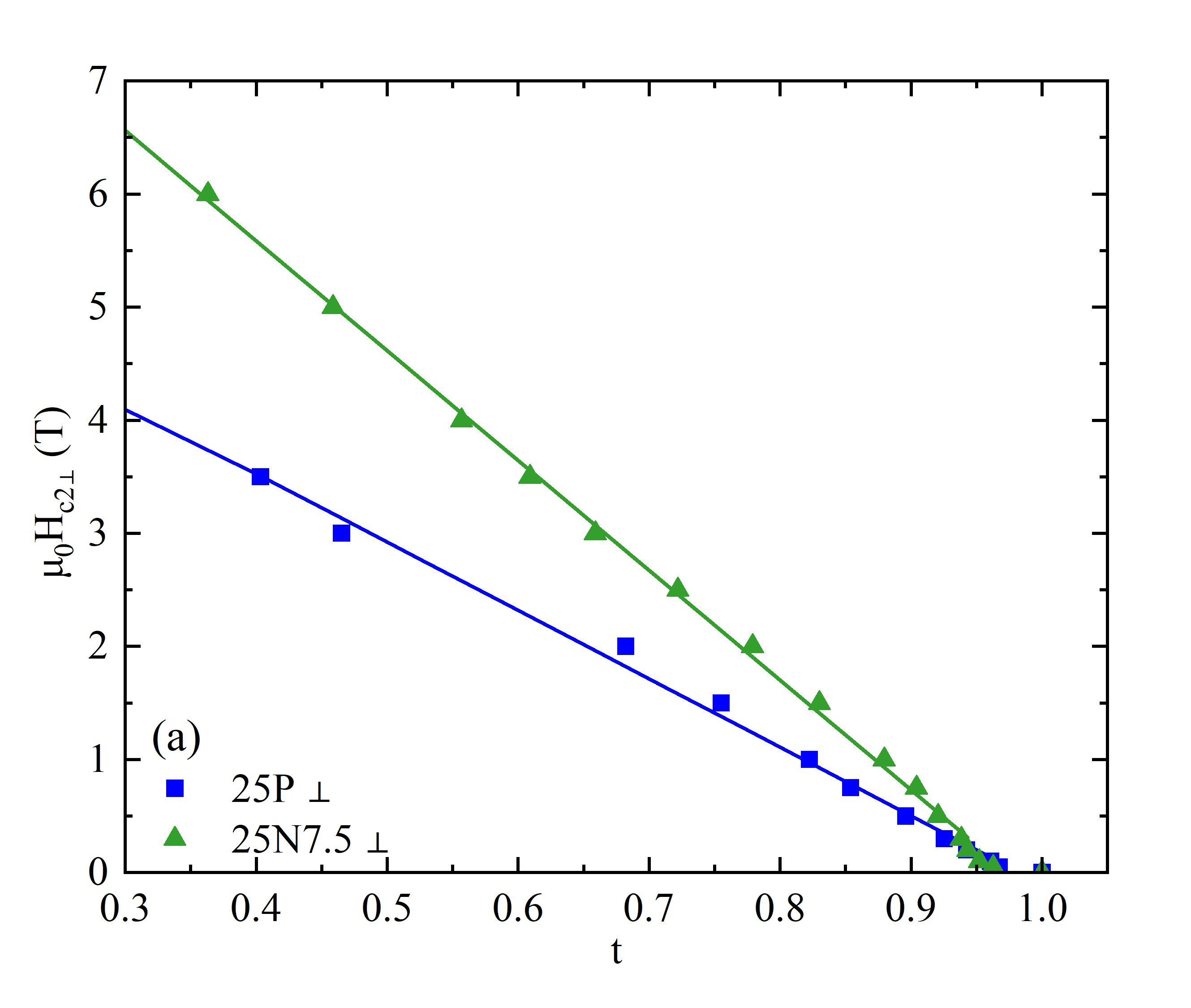}}}\\
    {\centerline{\includegraphics[width=9.0cm]{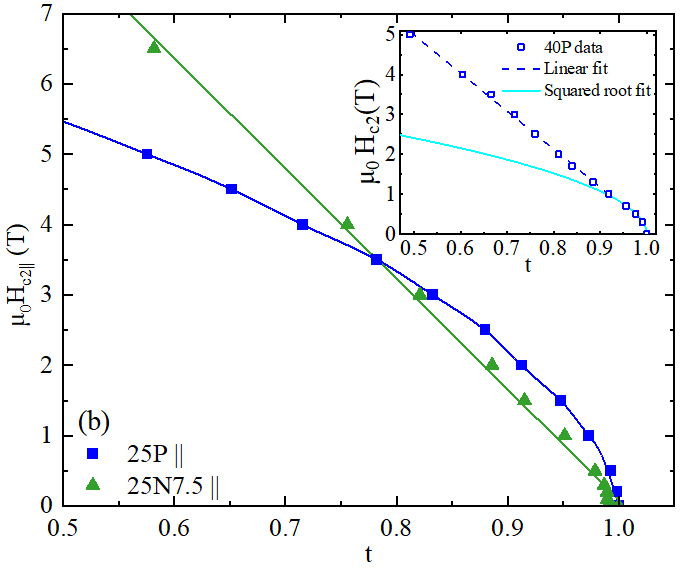}}}
    \caption{\justifying (a) $\mu_0H_{c2,\perp}$ and (b) $\mu_0H_{c2,\parallel}$ as functions of the reduced temperature $t$ for films 25P and 25N7.5. The straight solid lines represent the fitting of the linear parts of the $H_{c2}$ diagrams for both samples. For $\mu_0H_{c2,\parallel}$ of 25P, the squared root fitting is also reported. $\mu_0H_{c2||}$ for sample 40P is reported in the inset of panel (b), where the solid line corresponds to the fit of the square-root region of the plot, and the dashed line is the fit of its linear region.
}

\label{Hc2}
\end{figure}

Fig.~\textcolor{blue}{\ref{Hc2}}(a) displays $\mu_0H_{c2\perp}$ for samples 25P and 25N7.5 as a function of the reduced temperature $t=T/T_c$. For both samples, $\mu_0H_{c2\perp}(t)$ exhibits a positive concavity (PC) for $T$ approaching $T_c$, which is generally explained by non-homogeneity and disorder~\cite{Gray1977,Antonov2020}. As 
$T$ moves away from $T_c$ toward lower temperatures, the field scales linearly with the $t$, which allows the derivation of some fundamental superconducting parameters of the films, which are all reported in Table~\ref{results}. In particular, $\mu_0H_{c2\perp}\left(0\right)$ for both 25P and 25N7.5 was obtained by the linear fitting $H_{c2\perp}\left(t\right)=\frac{\phi_0}{2\pi \xi^2_{||}\left(0\right)}\left(1-t\right)$, where $H_{c2\perp}(0)=\frac{\phi_0}{2\pi \xi^2_{||}\left(0\right)}$ was left as fitting parameter. From the values of $H_{c2\perp}(0)$, the in-plane coherence length at $t=0$~K, $\xi_{||}\left(0\right)$, of the two samples was obtained. Then, the quasiparticle diffusion coefficient D was calculated as $D=\left(4k_B/\pi e\right)\cdot\left(\mu_0dH_{c2}/dT|_{T=T_c}\right)^{-1}$~\cite{Guimpel1986}, where $k_B$ and $e$ are the Boltzmann constant and the electron charge, respectively. The values of $D$ are consistent with the results obtained for $\rho$, since larger diffusion coefficients correspond to lower resistivities. Moreover, $D$ was used to estimate the density of states at the Fermi level from the free-electron Einstein's relation $N(0)=\left(e^2\rho^\text{10K} D\right)^{-1}$~\cite{Bartolf2010}. Additionally, from $T_c$ and $\rho^\text{10K}$ the magnetic penetration depth at zero temperature was derived as $\lambda(0)=1.05\times10^{-3}({\rho^\text{10K} T_c})^{-0.5}$~\cite{Tinkham}. Sample 25N7.5 exhibits a larger $\lambda(0)$ compared to 25P, which is consistent with expectations, since a larger $\lambda(0)$ is associated with increased disorder~\cite{Dzero2015}. The lower critical field at zero temperature $\mu_0H_{c1}(0)$ can be obtained as $\mu_0H_{c1}(0)=\Phi_0\ln{\kappa}/4\pi\lambda^2(0)$, where $\kappa=\lambda/\xi$ is the GL parameter~\cite{Kes1983}. The superconducting energy gap at zero temperature has been evaluated from the dirty-limit relation $\Delta(0)=\hbar\rho^\text{10K}\cdot\left(\lambda^2(0)\pi\mu_0\right)^{-1}$~\cite{Bartolf2010}. For both samples, a ratio $2\Delta(0)/k_BT_c=3.5$ is obtained, which is the expected value of for BCS superconductors~\cite{Udomsamuthirun1996}.

Fig.~\textcolor{blue}{\ref{Hc2}}(b) shows the plot of $\mu_0H_{c2||}\left(t\right)$ for the 25P and 25N7.5 films. For sample 25N7.5, $\mu_0H_{c2||}$ scales linearly with $t$, implying that the film behaves as a 3D superconductor for the entire temperature range~\cite{Tinkham,Attanasio1998,Tesauro2005,Cirillo2009}. In this case, the expression $H_{c2||}\left(t\right)=\frac{\phi_0}{2\pi \xi_{||}\left(0\right)\xi_\perp\left(0\right)}\left(1-t\right)$ was used to fit the experimental data, leaving $H_{c2||}\left(0\right)=\frac{\phi_0}{2\pi \xi_{||}\left(0\right)\xi_\perp\left(0\right)}$ as a fitting parameter. The out-of-plane coherence length at zero temperature, $\xi_\perp(0)$, was then obtained from the expression of $H_{c2||}\left(0\right)$ above using the $\xi_{||}(0)$ previously calculated. The fitting is reported as a solid line in Fig.~\ref{Hc2}(b). On the other hand, $\mu_0H_{c2||}\left(t\right)$ for 25P shows a squared-root-like trend on the entire temperature range. This result is typical of films of reduced thickness, and indicates a 2D behavior, which occurs when $\xi_\perp(T)$ is larger than $d$. In this case, the experimental data in all the $T$ range were fitted using the expression $H_{c2||}\left(t\right)=H_{c2||}(0)\left(1-t\right)^{0.5}$~\cite{Tinkham,Attanasio1998,Tesauro2005,Cirillo2009}, where $\mu_0H_{c2||}\left(0\right)=\frac{\sqrt{12}\phi_0}{2\pi \xi_{||}(0)d}$~\cite{Koorevaar} was left as the only fitting parameter. From the values of $\mu_0H_{c2||}\left(0\right)$ and $\xi_{||}(0)$, a thickness $d\sim6.5$~nm was calculated, which differs from the actual thickness of sample 25P. This thickness can be referred as an effective thickness ($d_\text{eff}$), and its meaning is discussed in Section IV. Since $\xi_\perp(T)$ decreases with temperature as $T$ goes to zero, there may be a certain crossover temperature $T^*$ so that, for $T<T^*$, $\xi_\perp(T) < d$ and the film behaves as a 3D superconductor. To study this effect, a thicker sample was tested. As expected, the $\mu_0H_{c2||}(T)$ measurements of sample 40P showed a 2D-3D crossover at $T^*=$~4.4~K ($t\simeq0.95$), as reported in the inset in Fig.~\ref{Hc2}(b). Consequently, $\mu_0H_{c2||}(T)$ scales as a root square above $T^*$ (2D) and linearly below $T^*$ (3D). Similarly to the case of sample 25P, from the root-squared fit of the 2D part of $\mu_0H_{c2||}(T)$, a $d_\text{eff}\sim20$~nm was obtained, using the $\xi_{||}(0)$ calculated from the 40P $\mu_0H_{c2\perp}(T)$ plot (not reported). {Not only does this suggest the presence of a $d_\text{eff}$ also in thicker film, but it also shows that $d_\text{eff}$ scales with the nominal film thickness. }

\subsection{Superconducting fluctuation effects above $T_c$}

The rounding effects on the in-plane electrical resistivity above the superconducting transition, measured in our samples, are analyzed in terms of thermodynamic fluctuations. 

{In disordered 2D superconductors, near the superconducting transition and above the critical temperature, quantum corrections to conductivity arise from weak localization (WL) and electron-electron interactions (EEI), and to excess conductivity due to superconducting fluctuations}~\cite{Altshuler1985,Dorin1993,Glatz2011}. {Both WL and EEI corrections give rise to an increase of the resistivity with decreasing temperature with a logarithmic temperature dependence. The complete expression for the total correction to the conductivity due to superconducting fluctuations of a disordered 2D superconductors in a perpendicular magnetic field, has been reported in the paper by Glatz et al.}~\cite{Glatz2011}. {In the Ginzburg-Landau region of fluctuations, i.e. close to $Tc$ and in zero magnetic field, the calculated total correction to conductivity gives the classical Aslamazov-Larkin (AL), Maki-Thompson (MT), and the density-of-state (DOS) contributions, and introduces the renormalization of the single-particle diffusion coefficient (DCR) term, which can be omitted close to $Tc$. In particular, close to $Tc$ and in zero magnetic field, the dominant correction due to superconducting fluctuations arises only from the AL and MT terms .}~\cite{Glatz2011}.

Within the classical theory of Aslamazov and Larkin~\cite{Aslamazov}, thermal fluctuations in a superconductor result in a finite probability of a Cooper pair formation above $T_c$, leading to an excess electrical conductivity. This effect depends on the system's dimensionality, and it is enhanced in thin superconducting films. Indeed, near $T_c$ the coherence length is generally larger than the film thickness and an agreement with two-dimensional superconductivity has been reported~\cite{Aslamazov,Maki,Thompson,Skocpol,Larkin,Destraz,Zhang,Liu}. 

In two dimensions, the direct Aslamazov-Larkin contribution to  conductivity is given by $\Delta \sigma ^\mathrm{AL}_{2D}=\frac{e^2}{16\hbar d } \varepsilon^{-1}$, where  $\varepsilon=ln\left(T/T_c\right)\approx \left( T-T_c\right)/ T_c$ 
and $d$ is the thickness of the superconducting film. An indirect contribution due to the interaction between the pairs and the normal electrons has been also considered and calculated by Maki and Thompson ~\cite{Maki,Thompson}, so that the main contribution to the excess conductivity in two dimensions and in zero applied magnetic field is written as

\begin{equation}
\Delta \sigma ^\mathrm{AL}_{2D}+\Delta \sigma ^\mathrm{MT}_{2D}=\frac{e^2}{16\hbar d } \varepsilon^{-1}+\frac{e^2}{8\hbar d } \frac{1}{\varepsilon-\delta} ln\left ( \frac{\varepsilon}{\delta} \right )\, .
\label{G0AL+MT}
\end{equation}
where $\delta=\pi\hbar/8\tau_{\phi}k_BT$ is the pair breaking parameter, with $\tau_{\phi}$ being the Thompson dephasing time~\cite{Thompson,Reizer}.

\begin{figure}
\centering
\includegraphics[width=10.5 cm]{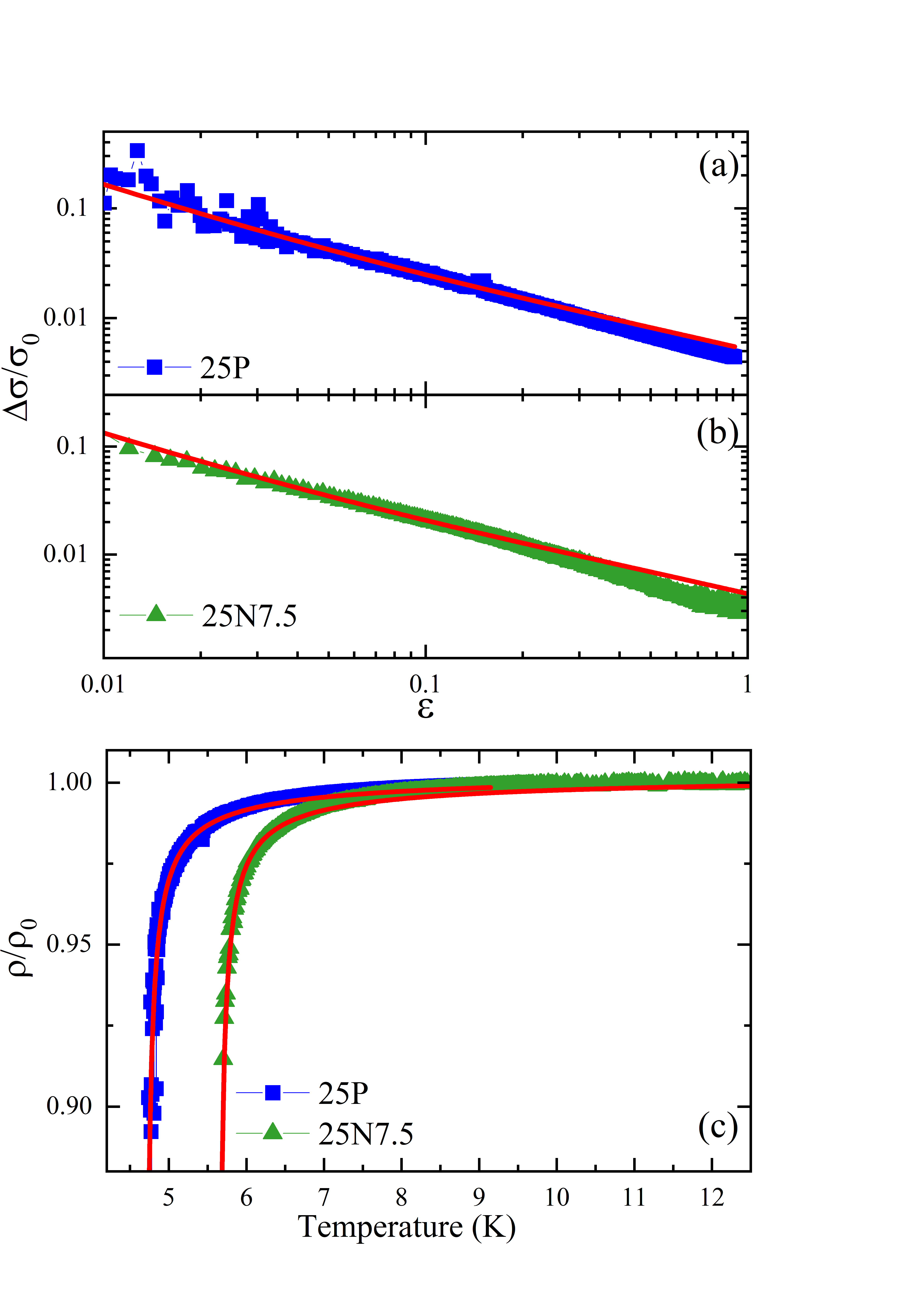}
\caption{\justifying
The normalized excess conductivity plotted as a function of the reduced temperature $\varepsilon$ in zero applied magnetic field for (a) 25P sample and (b) 25N7.5 sample. The red solid lines are the best fits to the data by equation~(\ref{G0AL+MT}). (c) The normalized resistivity  $\rho(T)/\rho^\text{10K}$  curves  for  the 25P and 25N7.5 samples in zero applied magnetic field and in the temperature range where the rouding due to thermodynamic fluctuations is observed.  The red solid lines are the best fits to the experimental data as described in the text.
}
\label{EXC-S25-N25}
\end{figure}

The normalized excess conductivity $\frac{\Delta\sigma(T)}{\sigma_0}$ was evaluated as $\frac{\Delta\sigma(T)}{\sigma_0}=\frac{\rho^\text{10K}}{\rho(T)}-\frac{\rho^\text{10K}}{\rho_n(T)}$, with $\rho(T)$ the sample resistivity, $\rho_n(T)$ the normal state resistivity and $\rho^\text{10K}$ the resistivity at $T=10$~K. In the investigated temperature range, $\rho_{n}(T)$ was assumed temperature-independent. In particular, $\rho_{n}(T)=85 \mu\Omega$~cm for sample 25P, and $\rho_{n}(T)=155 \mu\Omega$~cm for sample 25N7.5. {The samples 25P and 25N7.5 exhibit monotonic and weak temperature dependence with $RRR \approx 1$,
suggesting that the corrections from weak localization (WL) and electron-electron interaction (EEI) are small. This allows us to evaluate the dominant and
singular contributions of the excess conductivity by using the standard procedure of subtracting the normal state resistivity $\rho_n(T)$ as described above.} The results are reported in Fig.~\ref{EXC-S25-N25} (a) and (b) as a function of $\varepsilon$ for both 25P and 25N7.5 samples, respectively. In the temperature region where $\varepsilon \le 0.2$ the data of both samples can be described by a power law $\Delta\sigma(T)/\sigma_0 \sim \varepsilon^{-\alpha}$ with $\alpha\approx0.75$ suggesting the presence of non-negligible MT contribution $\Delta \sigma ^\mathrm{MT}_{2D}$. The experimental data of both samples were fitted using equation~(\ref{G0AL+MT}) with $d$ and $\delta$ as fitting parameters, and the results are reported as the red solid lines in Fig.~\ref{EXC-S25-N25} (a) and (b). For film 25P, the fit gives $d\sim8-9$~nm for the characteristic length of the 2D system, consistent with the $d_\text{eff}$ obtained from $\mu_0H_{c2||}(T)$, while, for film 25N7.5, $d\sim20$~nm, a value close to actual the thickness of the sample. Additionally, pair breaking parameter of $\delta=0.6$ and $\delta=0.5$ were obtained for 25P and 25N7.5 samples, respectively. Assuming $\delta\sim\pi\hbar/8\tau_{\phi}k_BT_c$, the phase breaking time $\tau_{\phi}$ close to $T_c$ is estimated to be $\approx 1$~ps for both samples. For comparison, reported values in the literature include $\tau_{\phi}=7$~ps for WSi~\cite{Zhang}, $\tau_{\phi}=2.5$~ps for NbN~\cite{Shinozakia}, and $\tau_{\phi}=4.5$~ps for Re$_{70}$W$_{30}$~\cite{Raffy1983}. Fig.~\ref{EXC-S25-N25} (c) shows the normalized resistivity  $\rho(T)/\rho^\text{10K}$ curves for the 25P and 25N7.5 samples at zero applied magnetic field and within the temperature range where the characteristic rounding due to thermodynamic fluctuations is observed.  The red solid lines in Fig.~\ref{EXC-S25-N25} (c) are the best fits to the experimental data by $\left(\rho^\text{10K}/\rho_n+\Delta \sigma/\sigma_0\right)^{-1}$ with $\Delta \sigma$ given by  equation~(\ref{G0AL+MT}) and the same values of the fitted parameters as in Fig.~\ref{EXC-S25-N25} (a) and (b). 

\section{Discussion and Conclusions}\label{Discussion}

Superconductivity in A15 tungsten is believed to be related to an amorphous phase, which constitutes a layer beneath the $\beta$ phase~\cite{Bagwe2024}. Although more comprehensive structural analyses are required, preliminary evidence indicates that similar conditions may be present also in our W$_{0.75}$Re$_{0.25}$. 

Comparing the XRD results in Fig.~\ref{XRD} with the $T_c$ values shown in Fig.~\textcolor{blue}{\ref{Tc_N}(a)}, a correlation between the crystallinity and the $T_c$ appears. In particular, the -P series, which exhibits both the $\alpha$ and the $\beta$ peaks, has the smallest $T_c$. On the contrary, $T_c$ increases for films of increasing disorder. The -N7.5 series, which has the larger $T_c$, appears almost amorphous, suggesting a correlation between crystallographic properties and superconductivity in our films. These results are supported by the analysis of the normal-state properties, such as RRR and $\rho$.

{While more in depth morphological characterizations are requested,} the behavior of the $\mu_0H_{c2||}(T)$ of the -P series suggest that the films may have a layered structure, with an amorphous layer of thickness $d_\text{eff}$ as the one responsible for the superconducting properties. {In spite of the phenomenological approach,} $d_\text{eff}$ is compatible with the $d$ found in the analysis of the superconducting fluctuation effects. This conclusion is also supported by the results of $\mu_0H_{c2||}(T)$ for the 25N7.5 film, which is mainly a single amorphous layer behaving as a 3D system in the presence of a magnetic field. Another possible indication of a layered structured is given by the PC of the $\mu_0H_{c2\,\perp}$ of both 25P and 25N7.5. In fact, a PC has also been observed in S/N multilayered films, as a result of the interactions at the S/N interfaces~\cite{Biagi1985,Takahashi1986,Fominov2001,Cirillo2003}. 

\setlength{\tabcolsep}{5pt}
\renewcommand{\arraystretch}{1.5}
\begin{table}
\vspace{3mm}
\begin{centering}
\caption{Characteristic properties of the samples 25P and 25N7.5.}
\label{results}
\begin{tabular}{c  c  c  c}
\hline
\hline
$ $ & 25P & 25N7.5 \tabularnewline
\hline 
$T_c$ (K) & 4.55 & 5.71 \tabularnewline

$\mu_0\,H_{c1\perp}(0)$ (mT) & 6.5 & 5.0\tabularnewline

$\mu_0\,H_{c2\perp}(0)$ (T) & 5.8 & 9.9 \tabularnewline

$\mu_0\,H_{c2||}(0)$ (T)& 7.4  & 14.9 \tabularnewline

$\xi_{\perp}(0)$ (nm)& - & 3.9 \tabularnewline

$\xi_{||}(0)$ (nm)& 7.6  & 5.8 \tabularnewline

$d_{\text{eff}}$ (nm)& 5.9 & - \tabularnewline

$\lambda(0)$ (nm) & {455} & {545}\tabularnewline

$D$ (m$^\text{2}$ s$^\text{-1}$) & {0.78$\cdot10^\text{-4}$} & {0.61$\cdot10^\text{-4}$}\tabularnewline

$N(0)$ (J$^\text{-1}$ m$^\text{-3}$) & {5.8$\cdot10^\text{47}$} & {4.1$\cdot10^\text{47}$}\tabularnewline

$\Delta(0)$ (meV) & {0.69} & {0.86}\tabularnewline

$2\Delta(0)/k_BT_c$ & {3.5} & {3.5}\tabularnewline

\hline  
\hline
\end{tabular}
\par\end{centering}
\end{table}

In conclusion, the transport and structural properties of superconducting W$_{0.75}$Re$_{0.25}$ thin films deposited by UHV DC magnetron sputtering in Ar and Ar/N$_2$ mixtures were investigated. We found that the transport properties strongly depends on the N$_2$ concentration during deposition, which significantly influences the crystalline phases. In absence of N$_2$, the coexistence of the $\alpha$ and the $\beta$ phase was observed, while the former was entirely suppressed in films grown in a proper Ar/N$_2$ mixture. Among our samples, the poorest crystallinity was found in the films deposited with N$_2$ being the 7.5\% of the total incoming flux. The superconducting properties are strongly correlated to the structure of the films. At fixed thickness, the largest values of $T_c$ and $\rho$ were observed in films deposited in a N$_2$ atmosphere. These films also exhibited effects associated with disorder and inhomogeneity. Our findings also suggest that the amorphous phase may form beneath the $\beta$ phase, as observed in pure W thin films~\cite{Bagwe2024}. This layered structure could have significant implications for understanding and optimizing the superconducting properties of W$_{0.75}$Re$_{0.25}$. Future research will focus on a more detailed analysis of the W$_{0.75}$Re$_{0.25}$ crystalline structure and its influence on superconductivity. I-V characteristics and critical vortex velocities, as well as magnoresistance will be explored on these films to assess their suitability for SNSPDs applications.

\end{document}


\title{Supplemental Material to\\Tuning Superconductivity in Sputtered W$_\textbf{0.75}$Re$_\textbf{0.25}$ Thin Films}
\author{F.~Colangelo}
\affiliation{Dipartimento di Fisica ``E.R. Caianiello'', Universit\`{a} degli Studi di Salerno, I-84084 Fisciano (Sa), Italy}
\affiliation{CNR-SPIN, c/o Universit\`{a} degli Studi di Salerno, I-84084 Fisciano (Sa), Italy}

\author{F.~Avitabile}
\affiliation{CNR-SPIN, c/o Universit\`{a} degli Studi di Salerno, I-84084 Fisciano (Sa), Italy}

\author{Z.~Makhdoumi~Kakhaki}
\affiliation{Dipartimento di Fisica ``E.R. Caianiello'', Universit\`{a} degli Studi di Salerno, I-84084 Fisciano (Sa), Italy}
\affiliation{CNR-SPIN, c/o Universit\`{a} degli Studi di Salerno, I-84084 Fisciano (Sa), Italy}

\author{A.~Kumar}
\affiliation{Dipartimento di Fisica ``E.R. Caianiello'', Universit\`{a} degli Studi di Salerno, I-84084 Fisciano (Sa), Italy}
\affiliation{CNR-SPIN, c/o Universit\`{a} degli Studi di Salerno, I-84084 Fisciano (Sa), Italy}

\author{A.~Di~Bernardo}
\affiliation{Dipartimento di Fisica ``E.R. Caianiello'', Universit\`{a} degli Studi di Salerno, I-84084 Fisciano (Sa), Italy}

\author{C.~Bernini}
\affiliation{CNR-SPIN Corso Perrone 24, I-16152 Genova, Italy}

\author{A.~Martinelli}
\affiliation{CNR-SPIN Corso Perrone 24, I-16152 Genova, Italy}

\author{A.~Nigro}
\affiliation{Dipartimento di Fisica ``E.R. Caianiello'', Universit\`{a} degli Studi di Salerno, I-84084 Fisciano (Sa), Italy}

\author{C.~Cirillo}
\affiliation{CNR-SPIN, c/o Universit\`{a} degli Studi di Salerno, I-84084 Fisciano (Sa), Italy}

\author{C.~Attanasio}
\affiliation{Dipartimento di Fisica ``E.R. Caianiello'', Universit\`{a} degli Studi di Salerno, I-84084 Fisciano (Sa), Italy}
\affiliation{CNR-SPIN, c/o Universit\`{a} degli Studi di Salerno, I-84084 Fisciano (Sa), Italy}
\affiliation{Centro NANO\_MATES, Universit\`{a} degli Studi di Salerno, I-84084 Fisciano (Sa), Italy}

\maketitle
\section{Simulations}
$R(T)$ measurements of superconducting W$_{0.75}$Re$_{0.25}$ thin films show a perpendicular-magnetic-field-activated peak effect (PE). The aim of this Supplemental Material is to numerically simulate the PE based on the proposed model and to provide a detailed description of how the simulation was carried out. Despite the attention that has been paid to carefully choosing the simulation parameters, the simulation is intended as a proof of the plausibility of the model rather than a complete demonstration.

\subsection{Introduction to the Simulation Problem}
The PE is most pronounced in the 25N7.5 film, attributed to its higher $T_c$, which enables $R(T)$ measurements up to $\mu_0 H_{\perp} = 6$~T. In contrast, the lower $T_c$ of the 25P film limits measurements to 3T, where the PE begins to emerge. For this reason, PE analysis focuses on the 25N7.5 sample, where it initially appears as a shoulder in $R(T)$ and later develops into distinct peaks.

It is well known that a PE may result because of the non-homogeneity of the samples and the use of the van der Pauw (vdP) method~\cite{Vaglio1993}. Let $R1$ and $R3$ be the resistances between the $I-I$ (current - current) and $V-V$ (voltage - voltage) contacts, respectively, while let $R2$ and $R4$ be those between the $I-V$ (current - voltage) contacts, as in the scheme reported in the insert in Fig.~\ref{25N7.5}. If the $T_c$ of either $R2$ or $R4$ is lower than those of both $R1$ and $R3$, the vdP measured resistance $\left(R_m^\text{\tiny vdP}\right)$ displays a peak. The formation of this PE can easily be deduced from the formula for $R_m^\text{\tiny vdP}$~\cite{Vaglio1993}:
\begin{equation}\label{eq:vdp}
    R_m^\text{\tiny vdP}=\frac{R1\,R3}{R1+R2+R3+R4}
\end{equation}

However, this model was developed in the zero-field condition, while in the case of our sample the PE is field activated. Our hypothesis is that the $R1$, $R2$, $R3$, and $R4$ have different $\mu_0 H_{c2\perp}(T)$ dependence, and that above a certain field, the $T_c$ of $R2$ or $R4$ becomes lower than the $T_c$ of the other two resistances, leading to the PE described above. According to Ginzburg-Landau theory, if $R1$ has a higher $T_c$ but a lower $\mu_0 H_{c2}(0)$ than $R2$, then there will be an intersection point in their $\mu_0 H_{c2}(T)$ phase diagrams. That is because from G-L theory~\cite{Tinkham},
\begin{equation}
    H_{c2\perp}(T) = H_{c2\perp}(0) - \frac{H_{c2\perp}(0)}{T_c} T
\end{equation}
\begin{onehalfspace} Consequently, if $T_c^{R1}>T_c^{R2}$ and $H_{c2}^{R1}(0)<H_{c2}^{R2}(0)$, then $H_{c2\perp}^{R1}(T)$ and $H_{c2\perp}^{R2}(T)$ have an intersection at \end{onehalfspace}

\begin{equation}
    T^{\text{int}}=\frac{H_{c2}^{R2}(0)-H_{c2}^{R1}(0)}{H_{c2}^{R2}(0)\,T_c^{R1}-H_{c2}^{R1}(0)\,T_c^{R2}}T_c^{R1}T_c^{R2}
\end{equation}
and it can easily be proved that $0<T^\text{int}<T_c^{R2}$.
According to this model, when the $T_c$ of $R1$ is larger than the $T_c$ of $R2$, then $H < H_{c2}(T^{\text{int}})$. Above $H_{c2}(T^{\text{int}})$, $T_c$ of $R2$ becomes higher than that of $R1$, creating conditions similar to those described in Ref.~\cite{Vaglio1993}, resulting in the peak effect. Additionally, this model is supported by the absence of PE when the samples are measured in the inline contact configuration. 

\subsection{Transition Curve Simulation Parameters}
The first issue to address is the simulation of the transition curves of the different resistances of the sample. Here, $T_c$ is defined as the temperature at which the resistance is half of its value just before the transition ($R_0$), i.e. $R\left(T_{c}\right)=0.5\,R_0$, while $\Delta T_c\equiv T_{c}^{\text{\tiny90\%}}-T_{c}^{\text{\tiny10\%}}$, where $T_{c}^{\text{\tiny90\% }}$ and $T_{c}^{\text{\tiny10\%}}$ are the temperatures such that $R\left(T_{c}^{\text{\tiny90\%}}\right)=0.9 \,R_0$ and $R\left(T_{c}^{\text{\tiny10\%}}\right)=0.1 \,R_0
$.

The superconducting resistive transition can be naively modelled by the expression 
\begin{equation}\label{eq:tanh}
    R(T)=\frac{R_0}{2}\left[\tanh{\left(\frac{T-T_c}{\delta}\right)}+1\right]
\end{equation}
where $\delta$ is a real number related to $\Delta T_c$.

Equation~(\ref{eq:tanh}) is consistent with our definition of $T_c$. In fact,
\begin{equation}
\begin{aligned}
    R\left(T=T_{c}\right)&=\frac{R_0}{2}\left[\tanh{\left(\frac{T_c-T_c}{\delta}\right)}+1\right]=\\
    &=\frac{R_0}{2}\left[\tanh{(0)}+1\right]=\\
    &=\frac{R_0}{2}
\end{aligned}
\end{equation}
In order to write equation~(\ref{eq:tanh}) explicitly in terms of the transition width $\Delta T_c$, a relationship between $\delta$ and $\Delta T_c$ has to be found. Substituting $T_{c}^{\text{\tiny90\%}}$ and $T_{c}^{\text{\tiny10\%}}$ into equation~(\ref{eq:tanh}) leads to 
\begin{equation}
\begin{aligned}
&\begin{cases}
    \frac{R_0}{2}\left[\tanh{\left(\frac{T_{c}^{\text{\tiny90\%}}-T_c}{\delta}\right)} + 1\right] = 0.9 \,R_0,\\
    \frac{R_0}{2}\left[\tanh{\left(\frac{T_{c}^{\text{\tiny10\%}}-T_c}{\delta}\right)} + 1\right] = 0.1 \,R_0
\end{cases}\\
\Longleftrightarrow&\begin{cases}
    \tanh{\left(\frac{T_{c}^{\text{\tiny90\%}}-T_c}{\delta}\right)} = 0.8,\\
    \tanh{\left(\frac{T_{c}^{\text{\tiny10\%}}-T_c}{\delta}\right)} = -0.8
\end{cases}\\
\Longleftrightarrow&\begin{cases}
    T_{c}^{\text{\tiny90\%}}=T_c+\delta\tanh^{-1}{(0.8)},\\
    T_{c}^{\text{\tiny10\%}}=T_c+\delta\tanh^{-1}{(-0.8)}
\end{cases}
\end{aligned}
\end{equation}
and $\delta\left(\Delta T_c\right)$ can be obtained as:
\begin{equation}
\begin{aligned}
    \Delta T_c&=T_{c}^{\text{\tiny90\%}}-T_{c}^{\text{\tiny10\%}}=\\
    &=\delta\left[\tanh^{-1}{(0.8)}+\tanh^{-1}{(-0.8)}\right]=\\
    &=2\delta\tanh^{-1}{(0.8)}\approx\\
    &\approx 2.197 \, \delta\\
    \Longleftrightarrow \delta &\approx \frac{\Delta T_c}{2.197}.
\end{aligned}
\end{equation}
Having found how to insert both $T_c$ and $\Delta T_c$ into the simulated transition, there is still the question of how to evaluate them in the first place.

\subsection{Evaluation of $T_c$ and $\Delta T_c$}
Concerning the evaluation of $T_c$ and $\Delta T_c$, there is a significant difference among the four resistances. For simplicity, we assume $R1=R3$ and $R2=R4$, as our aim is to reproduce the trend of $R(T)$ rather than its exact values. 

For $R1$, the critical temperature $T_c$ was determined from the experimental data using the 50\% criterion. As the magnetic field increases, the PE distorts the transition curves, starting from the tail and progressively affecting higher resistance regions. Up to $\mu_0 H = 2.5$~T, the midpoint of the transition appears unaffected by the PE, allowing $T_c$ to be directly obtained from the experimental data. For higher magnetic fields, $T_c$ values were extrapolated using a linear fit to the data points acquired below $\mu_0 H = 2.5$~T. The width $\Delta T_c$ of $R1$ was determined as $T_c^{\text{\tiny 90\%}} - T_c^{\text{\tiny 10\%}}$, reflecting the range of the transition as observed in $R_m^{\text{\tiny vdP}}$, which decreases only when the transition of $R1$ (or $R3$) begins. This approach is valid as long as $T_c^{\text{\tiny 90\%}}$ remains above the temperature at which the PE occurs. This condition is satisfied at all the applied field, except at $\mu_0 H = 6$~T, where $T_c^{\text{\tiny 90\%}}$ falls within the region affected by the PE. In this case, $\Delta T_c$ for $R1$ was estimated by extrapolating the $\Delta T_c(H)$ trend observed at lower fields.

\begin{figure}
\centerline{\includegraphics[width=9.5cm]{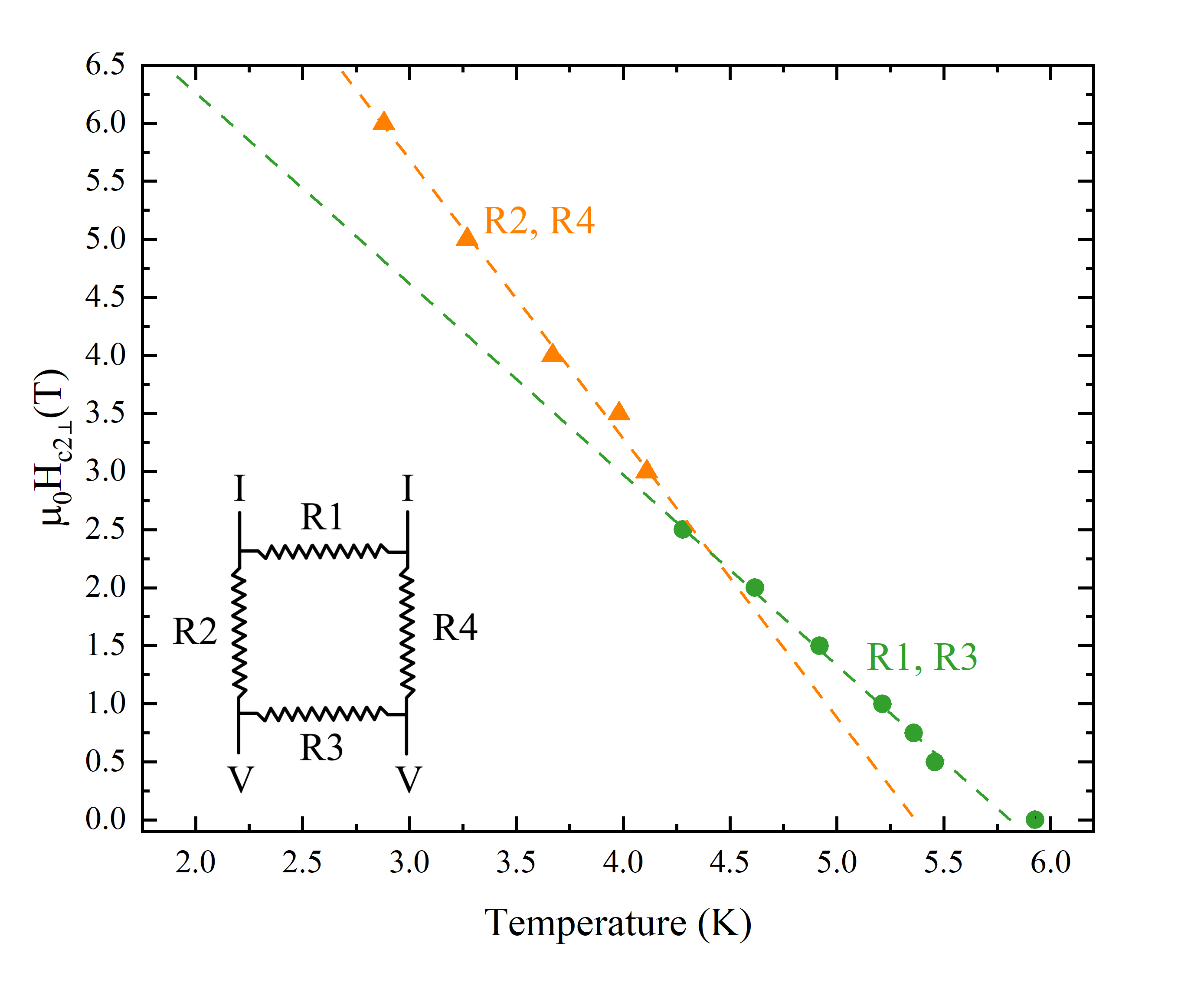}}
\caption{$\mu_0 H_{c2\perp}(T)$ phase diagram for the resistances $R1$ and $R3$ (green dots), and $R2$ and $R4$ (orange triangles). The points of $R1$ and $R3$ were directly obtained from the experimental $R(T)$ data, while those of $R2$ and $R4$ were derived from equation~(\ref{Tc_R2}). The scheme of the vdP four-resistance model is reported in the insert.} 
\label{25N7.5}
\end{figure}

According to equation~(\ref{eq:vdp}), the superconducting transition occurs at the $T_c$ of $R1$ (and $R3$). On the other hand, the superconducting transition of $R2$ (and $R4$) is reflected by an increase in the measured resistance, therefore the PE. For this reason, the temperature at which the PE occurs ($T_\text{peak}$) is related to the $T_c$ of $R2$. Additionally, as the peaks appear inside the transition region, a relative minimum point appears along with the peak in the $R(T)$ at a certain temperature $T_\text{min}>T_\text{peak}$. To obtain the relation among these quantities, many simulations with different $T_c$, $H$, and $\Delta T_c$ values were run. Notably, the simulations successfully reproduced the field-driven PE, even with the non-optimized $T_c$ of $R2$. In particular, the simulations revealed that for all the tested parameters, 
\begin{equation}\label{Tc_R2}
    T_c\approx0.465\left(T_\text{peak}+T_\text{min}\right).
\end{equation}
Additionally, having found the $T_c$ of both $R1$ and $R2$, it is possible to plot the $\mu_0 H_{c2\perp}(T)$ phase diagram for the two resistances. The graph is reported in Fig.~\ref{25N7.5}, and it shows a good match with our model. In fact, the trends of both $R1$ and $R2$ are linear, and an intersection at $\mu_0 H=2.5$~T is observed, which is the field value at which the slope change occurs in the $\mu_0 H_{c2\perp}(T)$ plot of the 25N7.5 sample. Notably, this intersection point emerged naturally from the simulations and was not externally introduced into the model.

The $\Delta T_c$ for $R2$ cannot be directly extracted from the experimental data. In fact, the peak widths depend on $\Delta T_c$ of both $R2$ and $R1$, as the curve's descent after the peak coincides with the transition of $R1$, and no method to decouple the two contributions was identified. Consequently, $\Delta T_c$ for $R2$ was treated as a free parameter to adjust the shapes of the simulated curves, without affecting the reliability of the simulations.

\begin{figure}
	\centering
	\subfloat
	{\includegraphics[width=.47\textwidth]{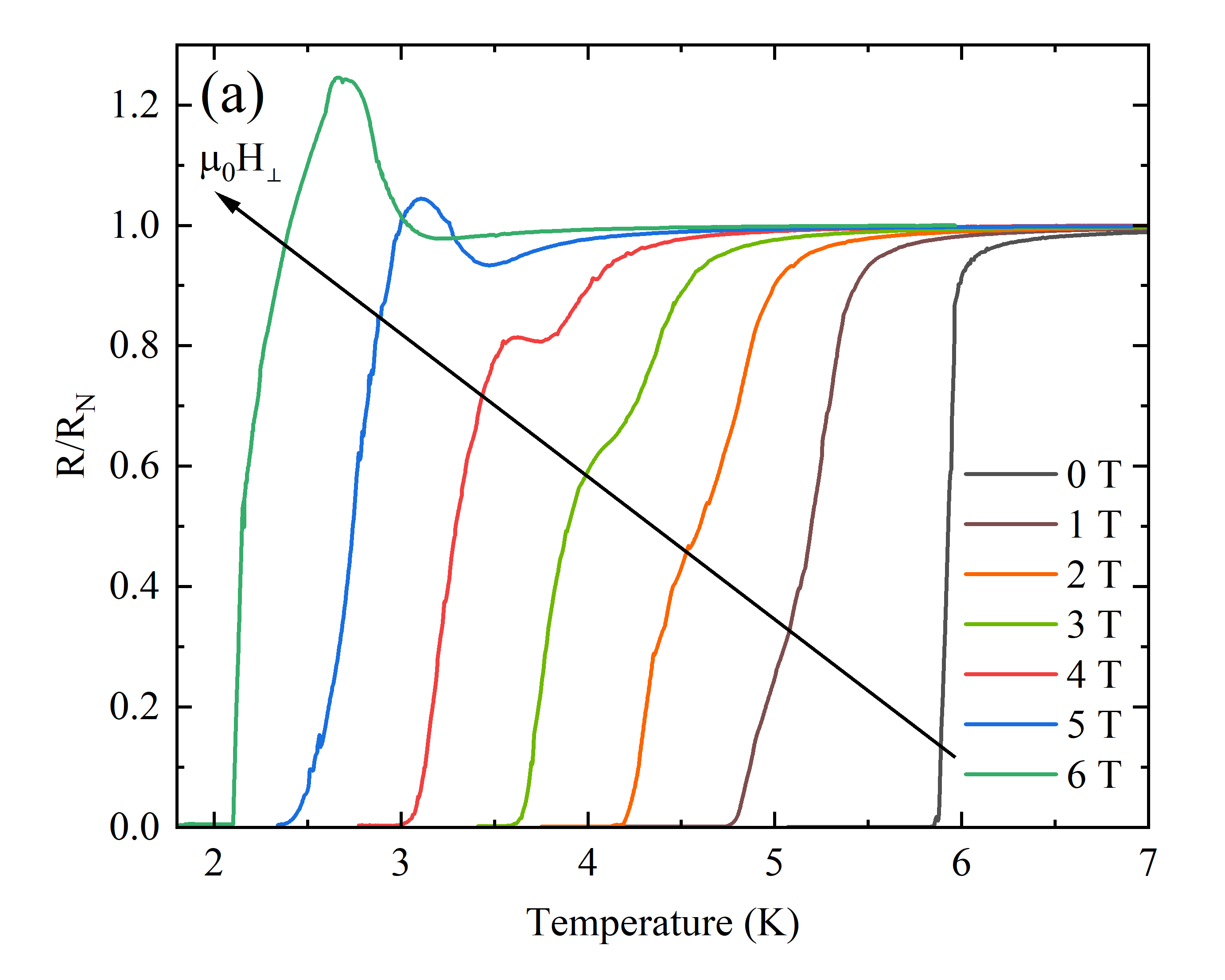}}$\;$
	{\includegraphics[width=.47\textwidth]{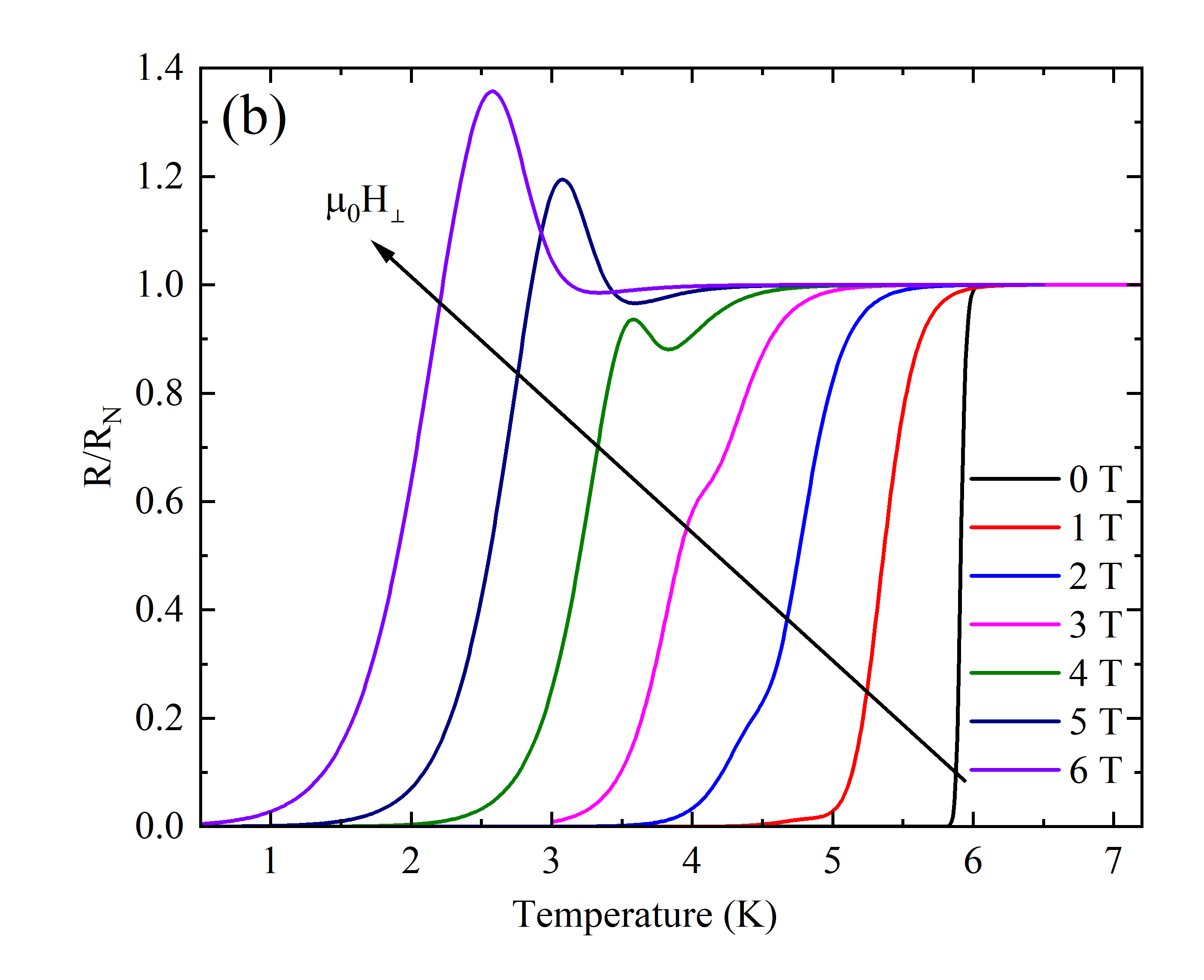}}
        \caption{(a) Experimental and (b) simulated normalized $R(T)$ curves of 25N7.5 in the vdP contacts configuration with a magnetic field perpendicularly applied to the film.}
	\label{vdP_comp_perp}
\end{figure}

\subsection{Simulation Results}
After having acquired all the parameters, simulations were performed. The results align well with the experimental data, and a comparison is shown in Fig.~\ref{vdP_comp_perp}. In facts, no PE is observed below 2.5~T, while, as the magnetic field increases above 2.5~T, a PE starts appearing in the simulated $R_m^\text{\tiny vdP}$. The PE starts as a shoulder in the lower part of the transition curve, and further increases in the field lead to an amplification of the shoulder and, eventually, the formation of distinct peaks. 

\begin{figure}
	\centering
	{\includegraphics[width=.48\textwidth]{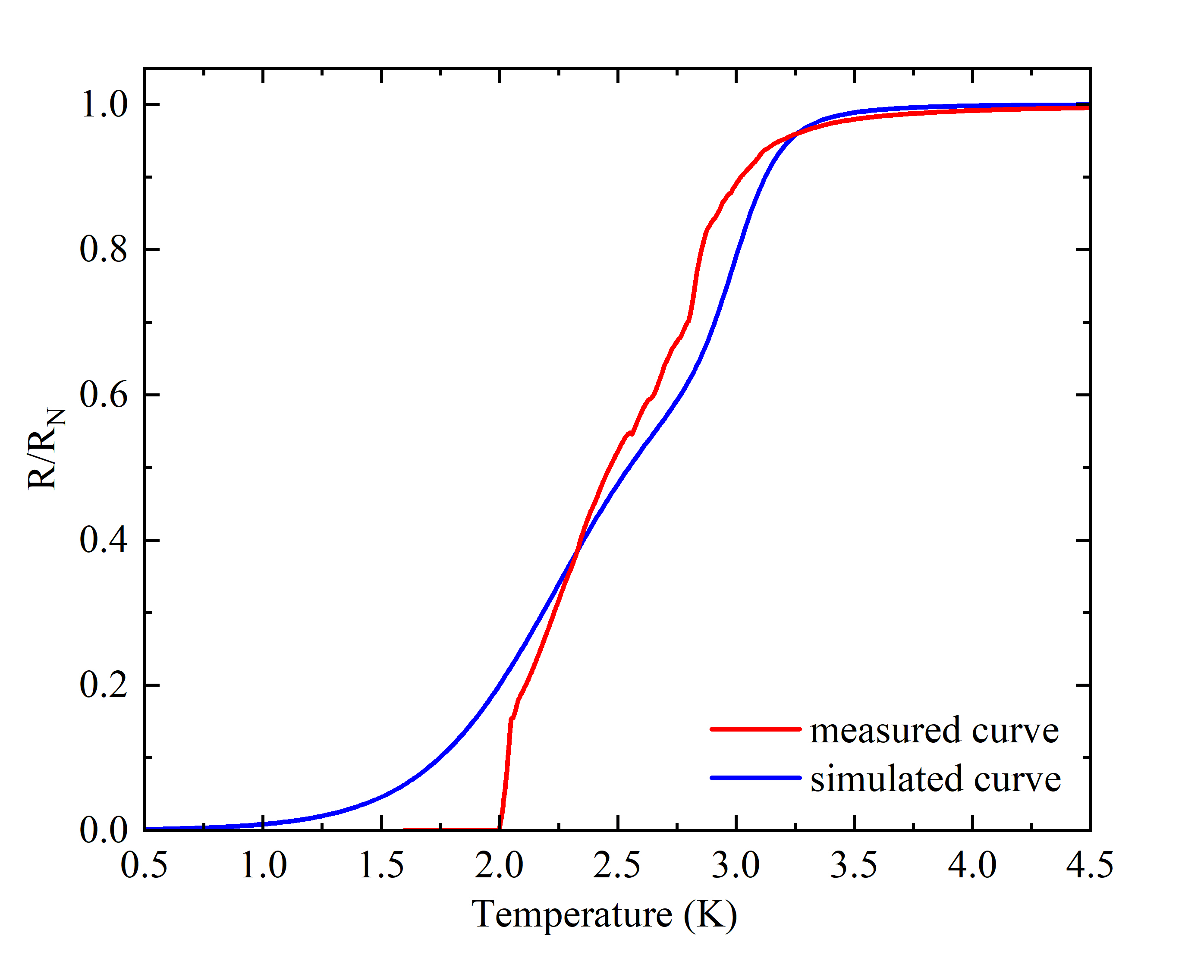}}
	  \caption{Measured and simulated normalized $R(T)$ curves of 25N7.5 with inline contacts at $\mu_0 H=6$~T, by solid red and blue line, respectively.}
	\label{inline}
\end{figure}

{These simulations indicate that the PE can be ascribed to the contact configuration due to the sample inhomogeneity. To further confirm this result, we measured the same film by using an in-line contact configuration, which was also simulated by the model. When the contacts are put in line the resistance} $\left(R_m^\text{\tiny inline}\right)$ {can be written as}~\cite{Vaglio1993}
\begin{equation}\label{eq:inline}
    R_m^\text{\tiny inline}=\frac{R1\left(R2+R3+R4\right)}{R1+R2+R3+R4}.
\end{equation}
To simulate $R_m^\text{\tiny inline}$, equation~(\ref{eq:inline}), which does not reproduce any PE, replaces equation~(\ref{eq:vdp}), while keeping the same exact parameters of the vdP simulations. Those were performed for various applied field values $H_\perp$, and no PE was observed. A comparison between the curves at 6~T for both vdP and inline configurations is shown in Fig.~\ref{inline}.

Finally, the parallel magnetic field configuration was also simulated, and no PE was observed. This can be explained by the fact that $H_{c2||}$ is generally larger than $H_{c2\perp}$, requiring higher fields to observe a PE. For $R1$, experimental data were used with $\mu_0 H_{c2||}(0) = 14.9$~T. For $R2$, the absence of a PE made it impossible to directly determine $\mu_0 H_{c2||}(T)$. However, by fixing the applied field to $\mu_0 H = 6.5$~T and varying $H_{c2||}$ of $R2$, we found that no PE appears if $\mu_0 H_{c2||} < 21$~T. Since $H_{c2||}$ for $R2$ is likely below this threshold, the result agrees with experimental observations.

\section{-N10 Series Data}
\begin{figure}
	\centering
	{\includegraphics[width=1.0\textwidth]{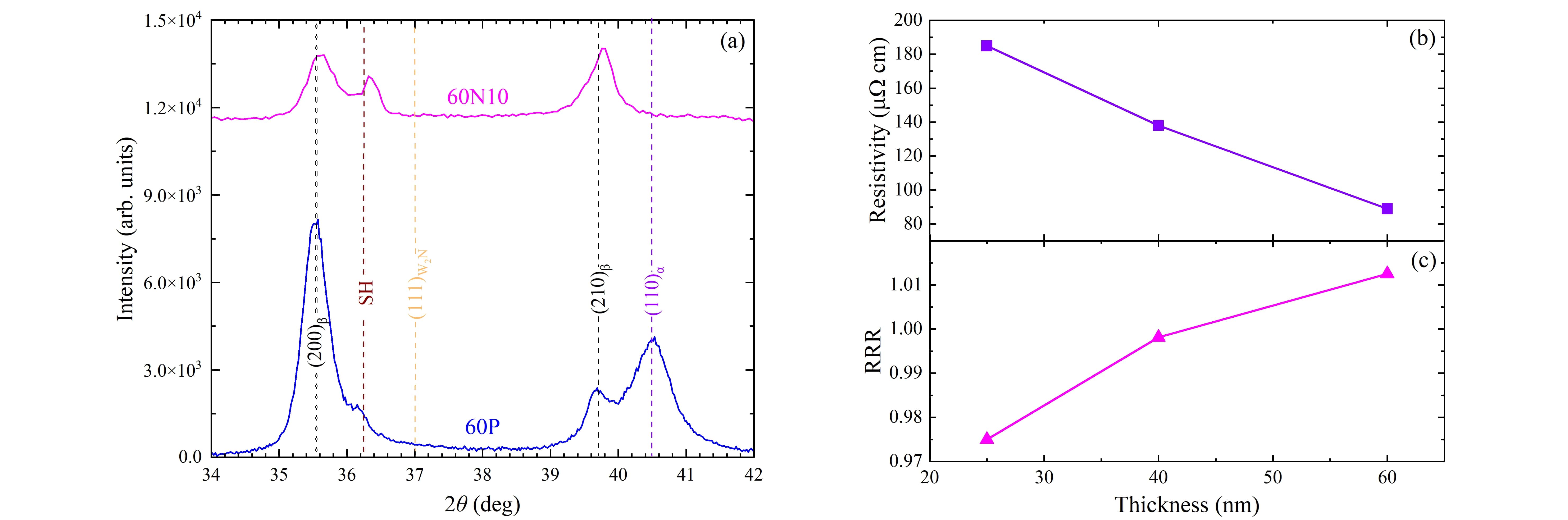}}
	  \caption{(a) XRD of samples 60N10 (pink solid line) compared to 60P (blue solid line); (b) resistivity and (c) RRR measurements of sample 60N10.}
	\label{N10}
\end{figure}

{Samples with 10\% N$_2$ (namely -N10 series) were also tested and characterized. 
As shown in Fig.}~\ref{N10}(a), {sample 60N10 (pink line) displays no peaks related to the $\alpha$-phase and those associated to the $\beta$-phase appear attenuated compared to sample 60P (in blue for reference), but not as much as for the -N7.5 series (see main text). $\rho$ and RRR as a function of the thickness are reported in Fig.}~\ref{N10}{(b) and (c), respectively. In both cases, the trends resemble what was observed for the -P series, rather than -N5 or -N7.5 series. This is coherent with the XRD results, which suggest a crystalline structure closer to the -P series. On the other hand, sample 25N10 exhibit $T_c=5.5$~K, which is comparable to the value obtained for sample 25N5.}

{While these results are interesting, they do not to match which the EDS measurements, which revealed no presence of N$_2$ in the -N10 series. For this reason, this sample series is discussed here, rather than in the main paper, in the hope that these results may be helpful for future studies.}

